\documentclass[pdflatex,sn-mathphys-num]{sn-jnl}% Math and Physical Sciences Numbered Reference Style
\usepackage{graphicx}%
\usepackage{multirow}%
\usepackage{amsmath,amssymb,amsfonts}%
\usepackage{amsthm}%
\usepackage{mathrsfs}%
\usepackage[title]{appendix}%
\usepackage{xcolor}%
\usepackage{textcomp}%
\usepackage{manyfoot}%
\usepackage{booktabs}%
\usepackage{algorithm}%
\usepackage{algorithmicx}%
\usepackage{algpseudocode}%
\usepackage{listings}%
\DeclareMathOperator{\sech}{sech}

\usepackage{mathtools}  % useful extensions to amsmath (recommended)
\usepackage{bm}         % bold math (\bm)
\usepackage{physics}    % convenient macros (optional)
\usepackage{hyperref}   % clickable links (optional)

\newcommand{\beqa}{\begin{eqnarray}}
\newcommand{\eeqa}{\end{eqnarray}}
\newcommand{\nn}{\nonumber}
\theoremstyle{thmstyleone}%
\theoremstyle{thmstyletwo}%

\theoremstyle{thmstylethree}%

\begin{document}

\title[Article Title]{A Self Propelled Vortex Dipole Model on a Surface of Variable Negative Curvature}

\author[1]{\fnm{Khushi} \sur{Banthia}}
\email{f20220982@hyderabad.bits-pilani.ac.in}

\author*[1]{\fnm{Rickmoy} \sur{Samanta}}
\email{rickmoy.samanta@hyderabad.bits-pilani.ac.in}

\affil[1]{\orgname{Birla Institute of Technology and Science, Pilani, Hyderabad Campus},
\postcode{500078}, \state{Telangana}, \country{India}}


\abstract{
We investigate vortex dipoles on surfaces of variable negative curvature, focusing on a catenoid of arbitrary throat radius $a$ as a concrete example. We construct the effective dynamical system by incorporating both inter-vortex interactions and the curvature-induced self-interaction arising from the geometry of the catenoid. The resulting Hamiltonian dynamics reveals that dipoles move along catenoid geodesics, consistent with recent works by Gustafsson (J. Nonlinear Sci. 32, 62 (2022)) and Drivas, Glukhovskiy, and Khesin (Int. Math. Res. Not. 2024, 10880–10894). We utilize the symplectic structure to construct a conserved momentum map $J$ associated with the $U(1)$ symmetry along the azimuthal direction. We explicitly demonstrate the conservation of both the Hamiltonian and $J$ for an arbitrary throat radius $a$. We then apply the formalism to analyze direct and exchange scattering of classical vortices on the catenoid surface. For comparison, we show that chiral pairs with similar initial configurations exhibit collective rotational motion (with azimuthal drift) instead of scattering. Finally, we construct a finite-dipole dynamical system on the catenoid based on dipole–dipole interactions and demonstrate the emergence of self-propulsion terms to leading order in dipole size. This provides a concrete realization, on a curved minimal surface, of the intuitive statement that a finite dipole \emph{propels} orthogonally to its axis, with a speed modulated by curvature.
}




\maketitle
\section{Introduction}

Vortex dipoles, consisting of two counter-rotating vortices of equal strength, are fundamental self-propelled structures that arise in many two-dimensional (2D) and quasi-2D flows. They represent the simplest mode of coherent transport and appear across a wide range of systems, from oceanic and atmospheric vortices to plasmas, superfluids, and Bose–Einstein condensates (BECs). In planar incompressible fluids, classical point-vortex theory \cite{aref,saffman,lin1,lin2} has been highly successful in describing their translational motion, pair interactions, and scattering dynamics. Finite-core and viscous effects account for dipole deformation, asymmetric decay, and interactions with walls or background shear \cite{Delbende2009,Dehtyriov2020,Flore1994,Hinds2009}. 

Many works have investigated the subtle role of curvature and topology of the underlying fluid domain in dipole and vortex motion \cite{bg,hally,kimok,kimura,newton1,newton2,crowdymarshall, Turner2010,Borisov2014,voigt, boattok, Koiller2009,boattod,Gustafsson2022,Lydon2022,khesin2024}, see also \cite{sam1,sam2,sam3}. When the underlying surface is non-Euclidean, the geometric coupling between curvature and vorticity modifies both the Hamiltonian structure and the resulting trajectories, together with a geometry-dependent self-interaction. Several recent studies \cite{sam1,sam2,sam3} have demonstrated quasi-periodic and chaotic regimes depending on initial separation and curvature. Such analyses highlight that curvature acts as an effective external field for vortex motion, breaking the translational symmetry that guarantees uniform propagation in the plane. 

A substantial body of mathematical literature \cite{boattok,Koiller2009,boattod,Gustafsson2022,khesin2024} has provided analytic arguments supporting Kimura’s geodesic conjecture, which states that tightly bound opposite-sign vortex pairs follow geodesics of the surface. These works show that the conjecture holds for infinitesimally close vortex pairs but acquires curvature-dependent corrections at finite separation.

Dipoles have also received considerable attention in recent condensed-matter experiments, particularly in the work of Neely \textit{et al.} \cite{Neely2010}, who observed the controlled nucleation and long-lived propagation of quantized vortex dipoles in an oblate BEC. Freilich \textit{et al.} \cite{Freilich2010} subsequently developed experimental techniques to extract real-time dipole trajectories, establishing connections between microscopic Gross–Pitaevskii dynamics and macroscopic vortex-particle models. More recent theoretical and numerical investigations \cite{Rooney2011,Goodman2015,white2012,White2014,Stagg2016}, closely tied to trapped-condensate experiments, have further strengthened the connection between vortex dipole dynamics and experimentally realizable geometries.

In this work, we formulate a dynamical description of vortex dipoles in an incompressible and inviscid fluid domain of variable negative curvature, using the catenoid of arbitrary throat radius $a$ as a concrete analytic example. The geometric construction and the finite-dipole model developed here extend naturally to other negatively curved surfaces. Starting from the geometry-dependent hydrodynamic Green’s function on the catenoid, we derive the explicit equations of motion for interacting vortices, incorporating both mutual interactions and curvature-induced self-interaction terms. 

The resulting symplectic structure yields a conserved momentum map associated with the $U(1)$ azimuthal symmetry, and we verify numerically the conservation of both the Hamiltonian and this momentum for an arbitrary throat radius. We find that tightly bound dipoles propagate along geodesics of the catenoid, providing explicit confirmation of Kimura’s geodesic conjecture and its recent generalizations to surfaces of variable curvature. We then demonstrate direct and exchange scattering of these dipoles on the catenoid. In contrast, co-rotating vortex configurations exhibit collective rotational states with azimuthal drift.

Extending the analysis to finite-sized dipoles, we construct an effective dynamical system that generalizes planar dipole models to minimal surfaces of varying negative curvature, yielding analytic expressions for curvature-corrected self-propulsion and orientation dynamics, which are explicitly validated numerically. Our work establishes the catenoid as a minimal surface suitable for analytic studies and provides a geometric framework for future investigations of self-propelled defects, vortex dipoles, and vortex clusters (along the lines of Ref.~\cite{sam3}) on curved manifolds.

Before proceeding to the organization of the paper, we comment on the relation to prior work and clarify our new contributions. Gustafsson~\cite{Gustafsson2022} shows that, in the vanishing-separation limit, a vortex dipole follows a geodesic on a general surface. Drivas, Glukhovskiy, and Khesin~\cite{khesin2024} rigorously generalize this result to asymmetric pairs, proving that singular vortex pairs follow magnetic geodesics, which reduce to ordinary geodesics in the dipole case. 

In contrast, we provide an explicit and fully dynamical analysis on the catenoid with arbitrary throat radius $a$. We derive the closed Hamiltonian system in global coordinates, construct and classify the catenoid geodesics analytically, and validate the dipole–geodesic correspondence across regimes. Going beyond the singular limit, we develop a finite-dipole reduced model with curvature-modulated self-propulsion and parallel-transport rotation terms, and demonstrate interacting multi-dipole phenomena, including direct and exchange scattering and collective rotation of co-rotating configurations.

The vortex dipoles considered here arise in two-dimensional incompressible ideal fluids governed by the Euler equations on curved surfaces, where point vortices provide a Hamiltonian reduction of the full flow. The same effective dynamics also describes quantized vortices in thin superfluid films and quasi-two-dimensional Bose–Einstein condensates confined to curved geometries, where vortex dipoles are experimentally realizable excitations. Our analysis therefore applies both to classical incompressible fluid layers and to superfluid systems in the hydrodynamic limit.

The catenoid is the simplest nontrivial minimal surface with negative Gaussian curvature and a tunable throat radius $a$, providing a geometrically controlled setting for studying vortex dynamics beyond constant-curvature manifolds. Its axial symmetry permits an explicit Hamiltonian formulation with a conserved momentum, while the analytic metric enables closed-form classification of geodesics into trans-throat, circular, and trapped families. Physically, the catenoid models thin classical or superfluid films supported on curved substrates or confined geometries, making it a natural laboratory for investigating curvature-modulated dipole propagation, scattering, and collective rotation.

We organize the paper as follows. In Sec.~\ref{ham} we present the Hamiltonian formulation and phase space, together with the associated momentum map corresponding to the azimuthal symmetry of the catenoid. In Sec.~\ref{geodesics} we compare vortex dipole motion with catenoid geodesics and provide numerical checks of the conservation laws for arbitrary throat radius. In Sec.~\ref{scattering} we demonstrate direct and exchange scattering of classical vortices on the catenoid surface. Section~\ref{cluster} analyzes collective rotational motion with azimuthal drift in co-rotating configurations. In Sec.~\ref{dipolesystem} we construct a reduced model for finite-sized dipoles and demonstrate the emergence of self-propulsion terms to leading order in dipole size, validated numerically. We conclude in Sec.~\ref{cncl}.
\section{Model Hamiltonian System and Phase Space}
\label{ham}

We begin by recalling the general Hamiltonian formulation of point-vortex motion on curved Riemannian surfaces, as developed in \cite{boattok, Koiller2009, boattod, Gustafsson2022, khesin2024}. While these works establish the geometric structure in full generality, our focus here is to specialize the theory to the \emph{catenoid} and derive explicit global expressions for the Hamiltonian, symplectic form, and conserved quantities for arbitrary throat radius $a$. In particular, we construct the closed Hamiltonian system governing $N$ vortices on the catenoid in global coordinates $(v,u)$, derive the explicit interaction Hamiltonian $H$, and compute the associated azimuthal momentum map $J$ arising from the rotational $U(1)$ symmetry. These concrete expressions form the foundation for the geodesic classification, dipole-limit reduction, and multi-dipole scattering analysis developed in the subsequent sections.

We consider the catenoid of throat radius $a>0$ described by
\[
X(v,u) = \big( a\cosh(v/a)\cos u,\; a\cosh(v/a)\sin u,\; v\big),
\qquad u\in[0,2\pi),\; v\in\mathbb{R}.
\]
This gives the metric
\[
g = \cosh^2\!\big(\tfrac{v}{a}\big)\,dv^2 
+ a^2\cosh^2\!\big(\tfrac{v}{a}\big)\,du^2
= \cosh^2\!\big(\tfrac{v}{a}\big)\,\big( dv^2 + a^2 du^2\big),
\]
with area element
\[
dA = a\cosh^2\!\big(\tfrac{v}{a}\big)\, dv\,du.
\]

Due to the periodicity of the azimuthal coordinate $u$, the hydrodynamic Green’s function may be written as
\[
G(v_i,u_i;v_j,u_j)
= \frac{1}{4\pi}\,\log\!\Big(
\cosh\!\big(\tfrac{v_i-v_j}{a}\big) - \cos(u_i-u_j)
\Big).
\]

For compactness, we define
\[
F_{ij} := \cosh\!\big(\tfrac{v_i-v_j}{a}\big) - \cos(u_i-u_j), 
\qquad 
h(v) := \cosh\!\big(\tfrac{v}{a}\big).
\]

For $N$ point vortices with strengths $\Gamma_i$ at positions $(v_i,u_i)$,
the Hamiltonian governing vortex interactions on the catenoid is
\beqa
H = \sum_{1\le i<j\le N} \Gamma_i\Gamma_j\, G(v_i,u_i;v_j,u_j)
-\frac{1}{4\pi}\sum_{i=1}^N \Gamma_i^2\,\log h(v_i). 
\label{hm}
\eeqa

The natural symplectic form induced by the area element is
\[
\omega = \sum_{i=1}^N \Gamma_i \, dA_i
= \sum_{i=1}^N \Gamma_i \, a\,h^2(v_i)\, dv_i\wedge du_i
= \sum_{i=1}^N \Gamma_i\, a\cosh^2\!\big(\tfrac{v_i}{a}\big)\, dv_i\wedge du_i.
\]

The Hamiltonian vector field 
\[
X_H=\sum_i(\dot v_i\partial_{v_i}+\dot u_i\partial_{u_i})
\]
satisfies $\iota_{X_H}\omega = dH$, which leads to Hamilton’s equations
\beqa
\Gamma_i\,a\,h^2(v_i)\,\dot v_i = \frac{\partial H}{\partial u_i}, 
\qquad
\Gamma_i\,a\,h^2(v_i)\,\dot u_i = -\frac{\partial H}{\partial v_i}.
\eeqa

Differentiating the pairwise Green function gives
\[
\frac{\partial G}{\partial u_i} 
= \frac{1}{4\pi}\,\frac{\sin(u_i-u_j)}{F_{ij}},
\qquad
\frac{\partial G}{\partial v_i} 
= \frac{1}{4\pi}\,\frac{1}{a}\,
\frac{\sinh\!\big(\tfrac{v_i-v_j}{a}\big)}{F_{ij}},
\]
and the local self-interaction term yields
\[
\frac{\partial}{\partial v_i}
\Big(-\frac{1}{4\pi}\Gamma_i^2\log h(v_i)\Big)
= -\frac{1}{4\pi}\Gamma_i^2\,\frac{1}{a}\,
\tanh\!\big(\tfrac{v_i}{a}\big).
\]

This leads to the final dynamical system:

\begin{equation}
\begin{aligned}
    \dot{v}_{i} &= \frac{1}{4\pi a\, h^{2}(v_{i})}
    \sum_{\substack{j=1 \\ j \neq i}}^{N} 
    \Gamma_{j}\, \frac{\sin(u_{i}-u_{j})}{F_{ij}}, \\[1.2ex]
    \dot{u}_{i} &= -\frac{1}{4\pi a^{2} h^{2}(v_{i})}
    \sum_{\substack{j=1 \\ j \neq i}}^{N} 
    \Gamma_{j}\, \frac{\sinh\!\big(\tfrac{v_{i}-v_{j}}{a}\big)}{F_{ij}}
    + \frac{1}{4\pi a^{2} h^{2}(v_{i})}\,
    \Gamma_{i}\, \tanh\!\big(\tfrac{v_{i}}{a}\big)
\end{aligned}
\label{dyneq}
\end{equation}

where
\[
h(v) = \cosh\!\left(\tfrac{v}{a}\right), 
\qquad 
F_{ij} = \cosh\!\left(\tfrac{v_{i}-v_{j}}{a}\right) - \cos(u_{i}-u_{j}).
\]

The above equations govern all subsequent dynamical analysis.

Before analyzing the dynamics, we construct an important conserved quantity. The catenoid is invariant under the $U(1)$ symmetry corresponding to rotations $u\mapsto u+\theta$. The associated conserved quantity follows from the corresponding momentum map.

Given the symplectic form 
\[
\omega = \sum_{i=1}^N \Gamma_i\,a\,\cosh^2\!\big(\tfrac{v_i}{a}\big)\, dv_i\wedge du_i,
\]
we contract $\omega$ with the infinitesimal generator $\partial/\partial u$. This yields
\[
\iota_{\partial/\partial u}\,\omega
= \sum_{i=1}^N \Gamma_i\,a\,\cosh^2\!\big(\tfrac{v_i}{a}\big)\,dv_i
= d\!\left( \sum_{i=1}^N \Gamma_i\,S(v_i)\right),
\]
which implies
\beqa
S'(v) = a\cosh^2(v/a).
\eeqa

Hence,
\[
S(v) = \int^v a\cosh^2\!\big(\tfrac{s}{a}\big)\,ds
= \tfrac{a}{2}v + \tfrac{a^2}{4}\sinh\!\big(\tfrac{2v}{a}\big).
\]

Thus, the conserved momentum is
\beqa
J = \sum_{i=1}^N \Gamma_i\left(
\frac{a}{2}\,v_i + \frac{a^2}{4}\,\sinh\!\big(\tfrac{2v_i}{a}\big)
\right),
\qquad \frac{dJ}{dt}=0. 
\label{jdef}
\eeqa

Along with this, we also have conservation of the Hamiltonian $H$ defined in Eq.~\ref{hm}. The conservation of $J$ and $H$ will serve as important diagnostic tools for the subsequent analysis.
\section{Investigation of Geodesics}
\label{geodesics}

\begin{figure}[htbp!]
    \centering
    % ---- Row 1 ----
    \begin{tabular}{cc}
        \includegraphics[width=0.75\textwidth]{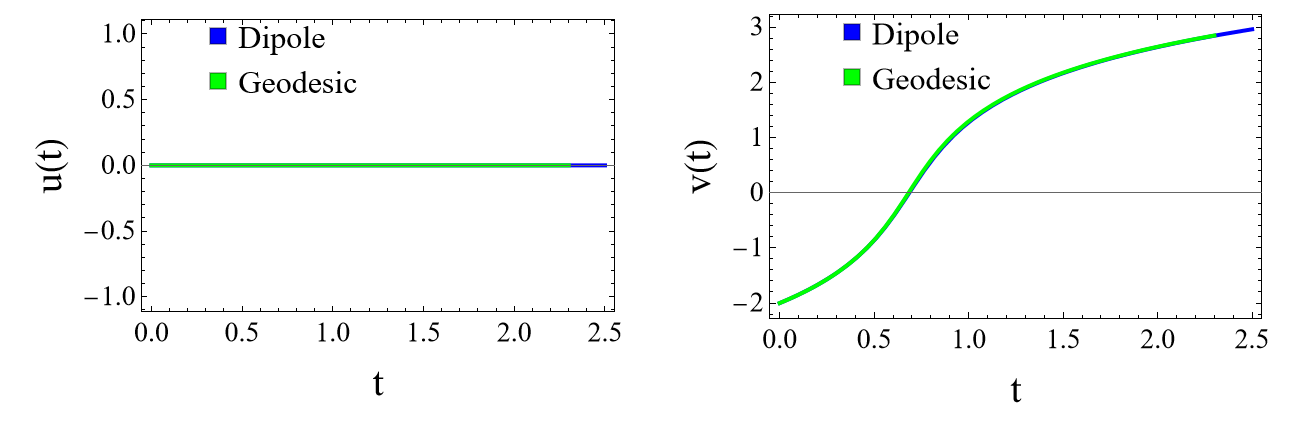} &
        \includegraphics[width=0.40\textwidth]{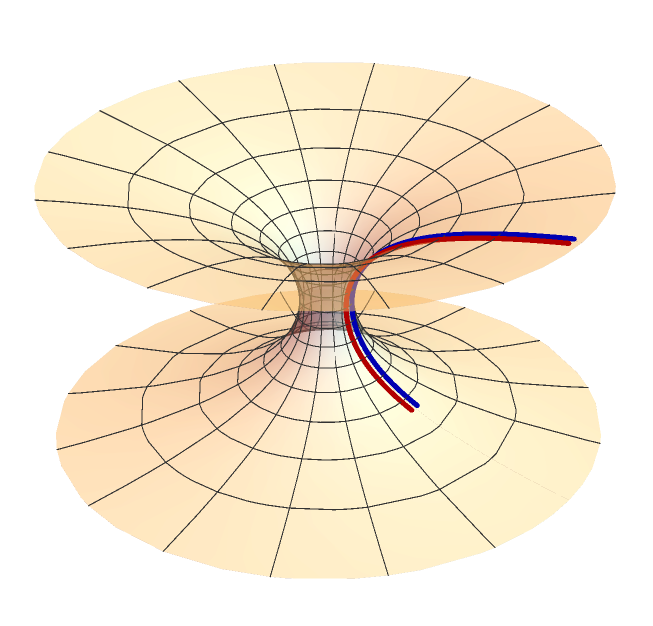} \\
        (a) $u(t)$ and $v(t)$ evolution & (b) Geodesic visualization \\
    \end{tabular}

    \vspace{0.5em}

    % ---- Row 2 ----
    \begin{tabular}{cc}
        \includegraphics[width=0.45\textwidth]{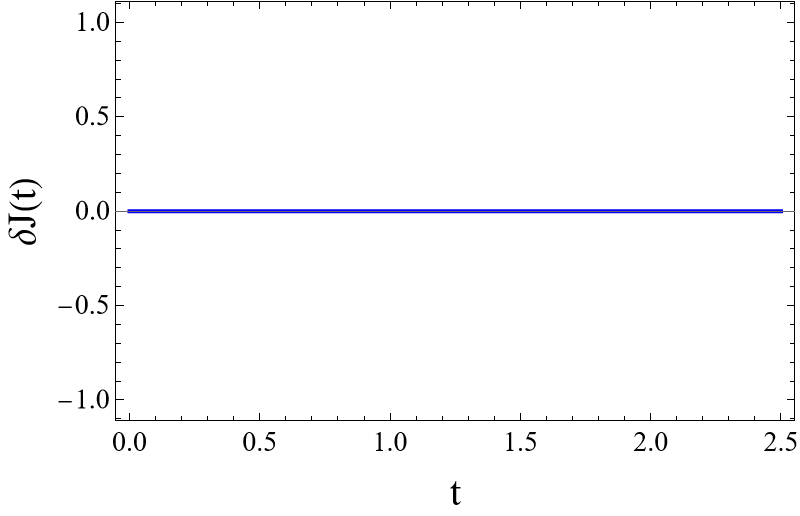} &
        \includegraphics[width=0.45\textwidth]{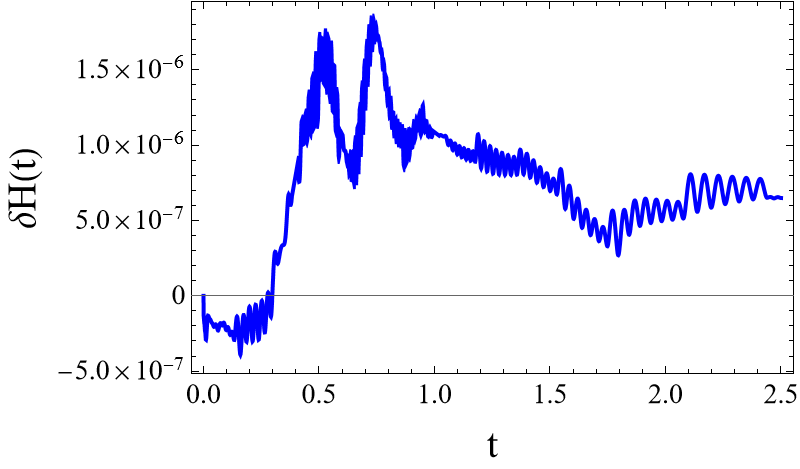} \\
        (c) $\delta J(t)$ vs $t$ & (d) $\delta H(t)$ vs $t$ \\
    \end{tabular}

    \caption{
        Meridional geodesic evolution with initial conditions $(u_1,v_1,u_2,v_2)=(\epsilon,-2,-\epsilon,-2)$, where $\epsilon=0.05$. This choice yields $J=0$ exactly and $\Lambda=0$ to leading order.
        Top row: time evolution and geodesic embedding.
        Bottom row: deviations of the conserved quantities $\delta J$ and $\delta H$.
        In the $(u,v)$ trajectory plots, the green curve denotes the analytical geodesic solution, while the blue curve shows the corresponding vortex--dipole evolution. For clarity, the geodesic curve is intentionally terminated at an earlier value of the affine parameter so that both trajectories can be visualized distinctly.
    }
    \label{figgdsc1}
\end{figure}

\begin{figure}[htbp!]
    \centering
    % ---- Row 1 ----
    \begin{tabular}{cc}
        \includegraphics[width=0.75\textwidth]{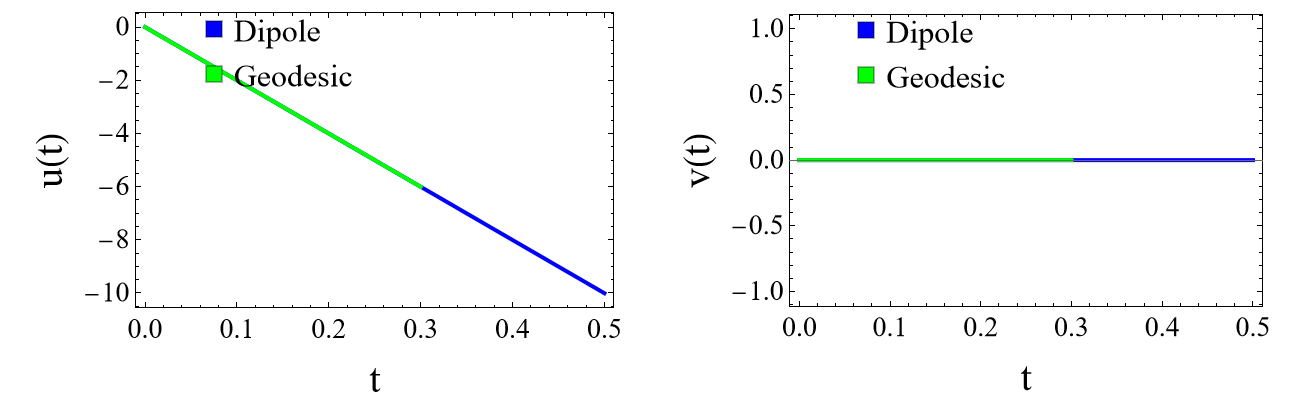} &
        \includegraphics[width=0.40\textwidth]{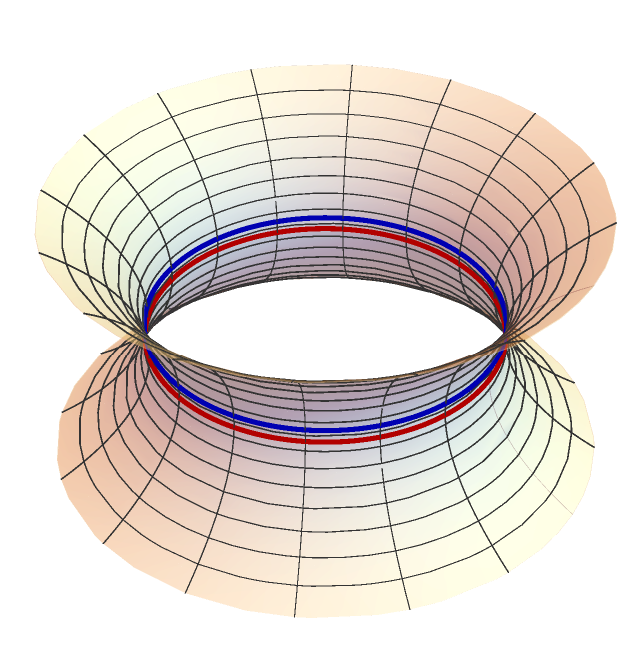} \\
        (a) $u(t)$ and $v(t)$ evolution & (b) Geodesic visualization \\
    \end{tabular}

    \vspace{0.5em}

    % ---- Row 2 ----
    \begin{tabular}{cc}
        \includegraphics[width=0.45\textwidth]{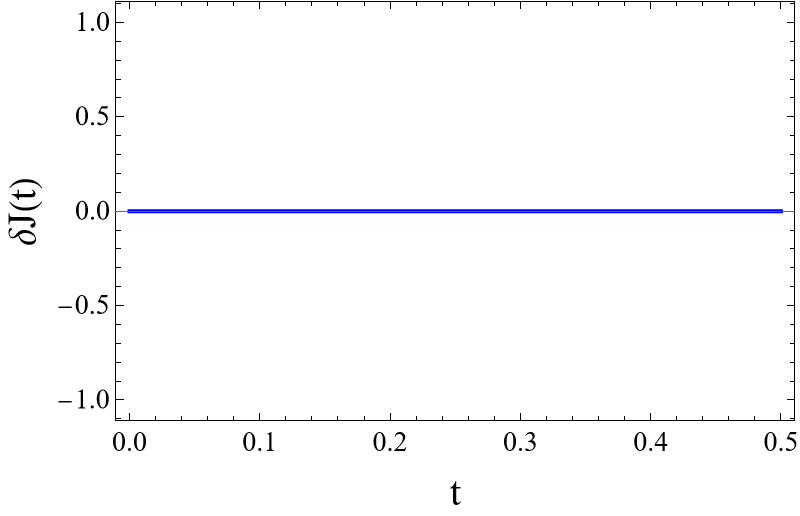} &
        \includegraphics[width=0.45\textwidth]{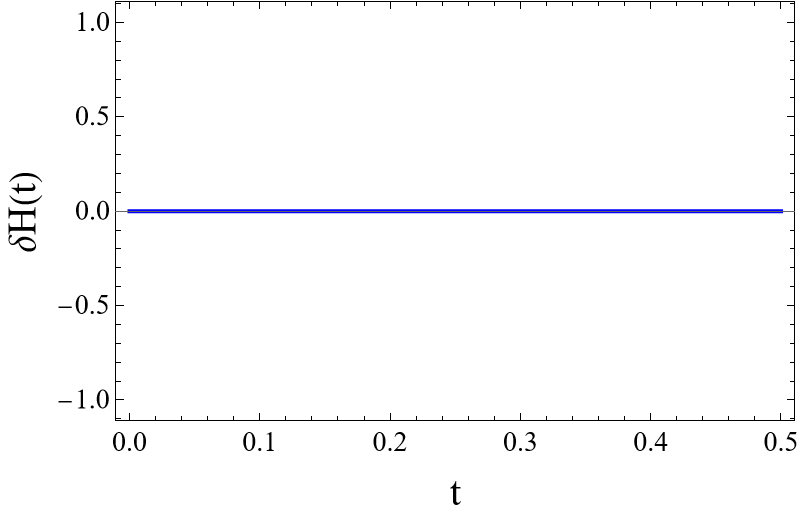} \\
        (c) $\delta J(t)$ vs $t$ & (d) $\delta H(t)$ vs $t$ \\
    \end{tabular}

    \caption{
        Critical (neck circle) geodesic evolution with initial conditions $(u_1,v_1,u_2,v_2)=(0,\epsilon,0,-\epsilon)$, where $\epsilon=0.05$. This choice yields $\Lambda=1$ to leading order.
        Top row: time evolution and geodesic embedding.
        Bottom row: deviations of the conserved quantities $\delta J$ and $\delta H$.
        In the $(u,v)$ trajectory plots, the green curve denotes the analytical geodesic solution, while the blue curve shows the corresponding vortex--dipole evolution. For clarity, the geodesic curve is intentionally terminated at an earlier value of the affine parameter so that both trajectories can be visualized distinctly.
    }
    \label{figgdsc2}
\end{figure}

\begin{figure}[htbp!]
    \centering
    % ---- Row 1 ----
    \begin{tabular}{cc}
        \includegraphics[width=0.75\textwidth]{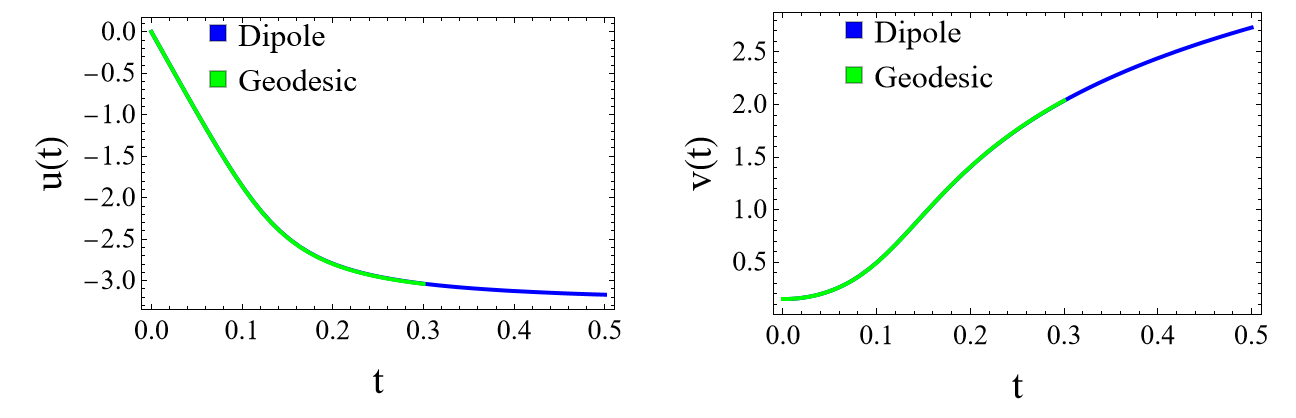} &
        \includegraphics[width=0.40\textwidth]{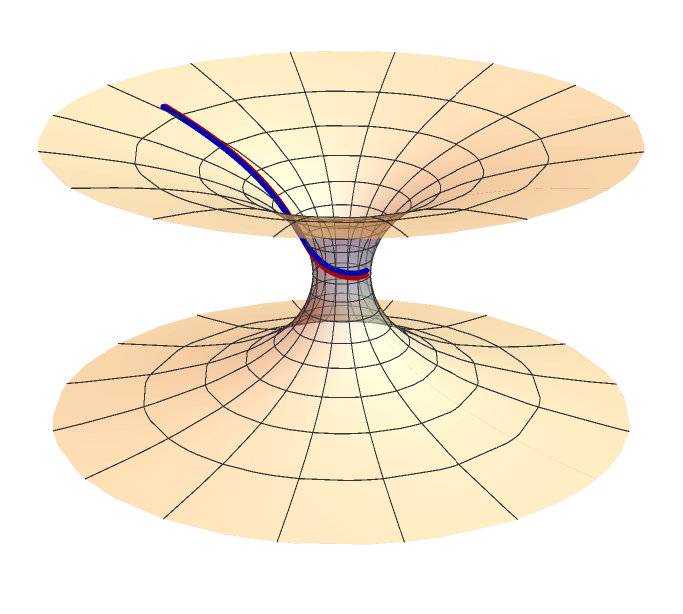} \\
        (a) $u(t)$ and $v(t)$ evolution & (b) Geodesic visualization \\
    \end{tabular}

    \vspace{0.5em}

    % ---- Row 2 ----
    \begin{tabular}{cc}
        \includegraphics[width=0.45\textwidth]{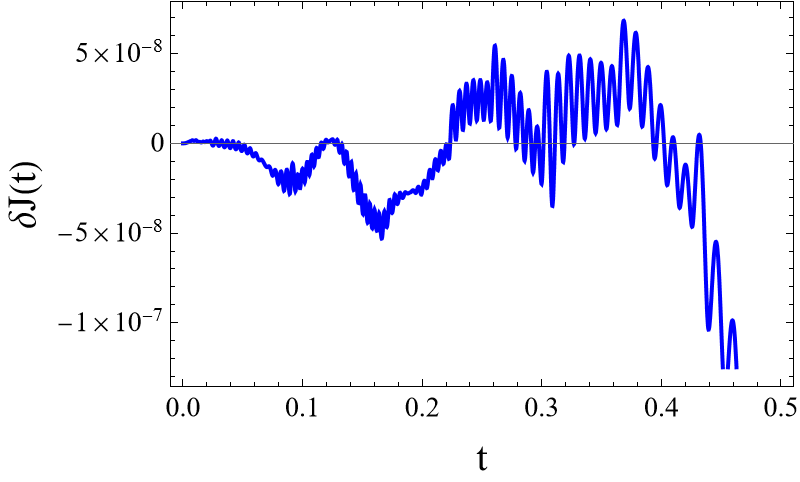} &
        \includegraphics[width=0.45\textwidth]{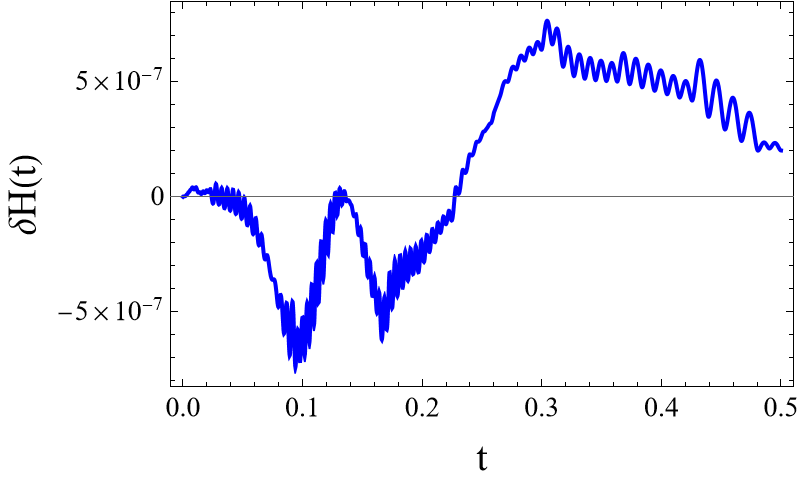} \\
        (c) $\delta J(t)$ vs $t$ & (d) $\delta H(t)$ vs $t$ \\
    \end{tabular}

    \caption{
        Trapped one-sided geodesic evolution with initial conditions $(u_1,v_1,u_2,v_2)=(0,0.15+\epsilon,0,0.15-\epsilon)$, where $\epsilon=0.05$. This choice yields $|\Lambda|>1$ to leading order.
        Top row: time evolution and geodesic embedding.
        Bottom row: deviations of the conserved quantities $\delta J$ and $\delta H$.
        In the $(u,v)$ trajectory plots, the green curve denotes the analytical geodesic solution, while the blue curve shows the corresponding vortex--dipole evolution. For clarity, the geodesic curve is intentionally terminated at an earlier value of the affine parameter so that both trajectories can be visualized distinctly.
    }
    \label{figgdsc3}
\end{figure}

\begin{figure}[htbp!]
    \centering
    % ---- Row 1 ----
    \begin{tabular}{cc}
        \includegraphics[width=0.75\textwidth]{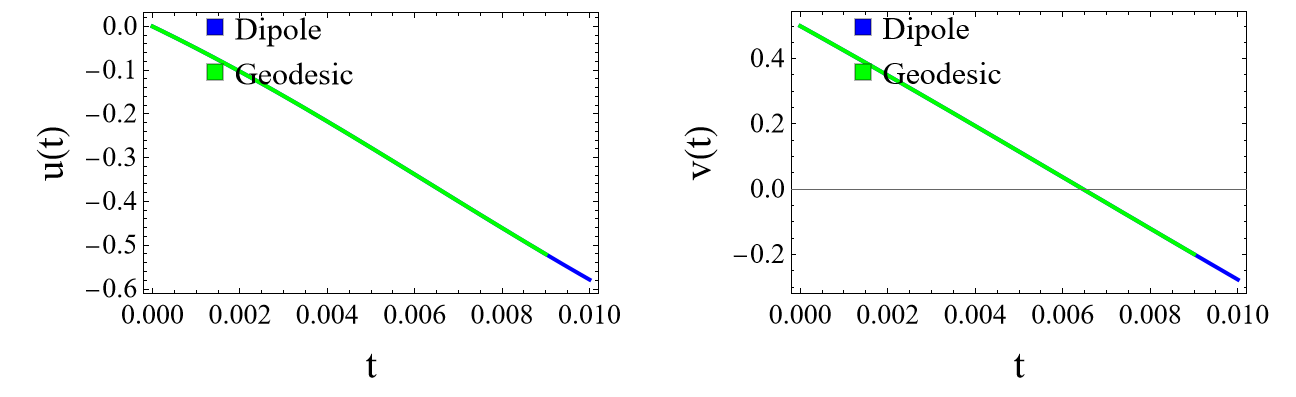} &
        \includegraphics[width=0.40\textwidth]{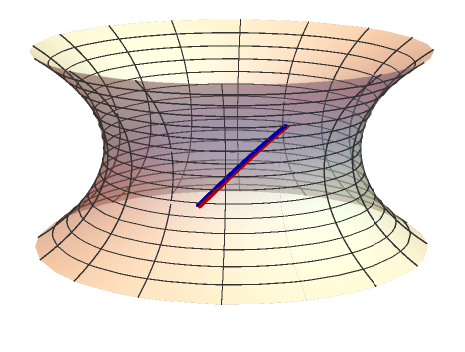} \\
        (a) $u(t)$ and $v(t)$ evolution & (b) Geodesic visualization \\
    \end{tabular}

    \vspace{0.5em}

    % ---- Row 2 ----
    \begin{tabular}{cc}
        \includegraphics[width=0.45\textwidth]{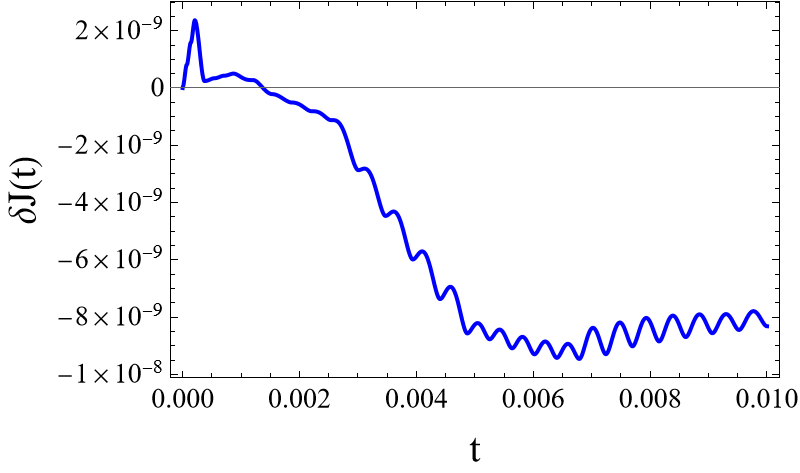} &
        \includegraphics[width=0.45\textwidth]{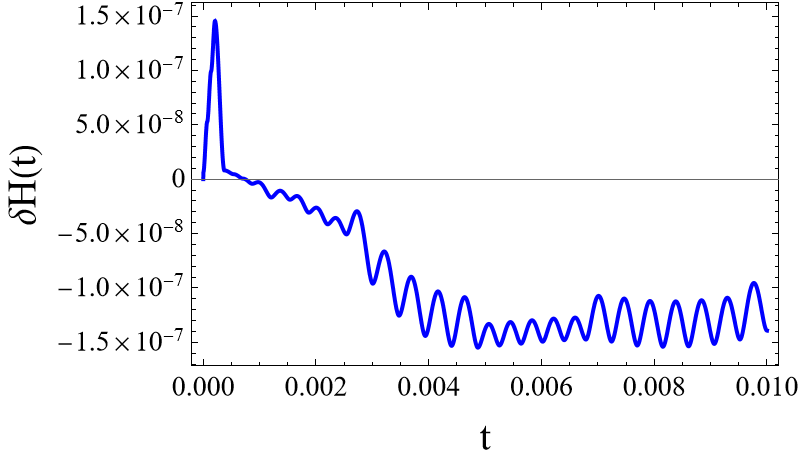} \\
        (c) $\delta J(t)$ vs $t$ & (d) $\delta H(t)$ vs $t$ \\
    \end{tabular}

    \caption{
       Subcritical geodesic evolution generated by a vortex dipole with initial
conditions $(u_1,v_1,u_2,v_2)=(-0.00744,\,0.50484,\,0.00744,\,0.49516)$,
corresponding to $\Lambda\simeq0.6$ with $|\Lambda|<1$.
Top row: time evolution and geodesic embedding.
Bottom row: deviations of the conserved quantities $\delta J$ and $\delta H$.
In the $(u,v)$ plane, the green curve shows the analytic geodesic, while
the blue curve denotes the vortex--dipole evolution.
    }
    \label{figgdsc4}
\end{figure}

In the study of point vortices on curved surfaces, the so-called \textit{Kimura conjecture} proposes that an infinitesimally close vortex pair of equal and opposite circulations (a vortex dipole) moves, in the vanishing-separation (``dipole'') limit, along a geodesic of the underlying surface metric. Kimura first formulated and verified this behavior on constant-curvature surfaces~\cite{kimura}. Subsequent rigorous work by Boatto and Koiller recast the vortex-pair Hamiltonian on a general Riemannian surface, providing a proof of Kimura’s conjecture by showing that the reduced dynamics converge to the geodesic equation~\cite{Koiller2009}. More recently, Gustafsson~\cite{Gustafsson2022} extended this analysis, deriving directly from the vortex Hamiltonian that, in the dipole limit, the centroid motion satisfies the geodesic equation (with a reparameterized time). The latest advance by Drivas, Glukhovskiy, and Khesin~\cite{khesin2024} generalized the result to asymmetric vortex pairs (nonzero total circulation), showing that such singular pairs follow \emph{magnetic geodesics}; in the special case of vanishing total circulation (a true dipole), one recovers Kimura’s original geodesic motion.

With the dynamical equations in Eq.~\eqref{dyneq} at hand, we now proceed to an explicit verification of this result on a catenoid of arbitrary throat radius. For comparison with the dipole trajectory, we construct the corresponding geodesic equations and their solutions. The nonvanishing Christoffel symbols are
\begin{equation}
\Gamma^{v}_{vv}=\frac{1}{a}\tanh\!\Bigl(\frac{v}{a}\Bigr), \qquad
\Gamma^{v}_{uu}=-a\,\tanh\!\Bigl(\frac{v}{a}\Bigr), \qquad
\Gamma^{u}_{uv}=\Gamma^{u}_{vu}=\frac{1}{a}\tanh\!\Bigl(\frac{v}{a}\Bigr).
\end{equation}
Hence, the geodesic equations
$\ddot x^{i}+\Gamma^{i}_{jk}\dot x^{j}\dot x^{k}=0$
take the explicit form
\begin{align}
\ddot v+\frac{1}{a}\tanh\!\Bigl(\frac{v}{a}\Bigr)\dot v^{2}
- a\,\tanh\!\Bigl(\frac{v}{a}\Bigr)\dot u^{2}&=0,
\label{eq:geo-v}\\[3pt]
\ddot u+\frac{2}{a}\tanh\!\Bigl(\frac{v}{a}\Bigr)\dot u\,\dot v&=0.
\label{eq:geo-u}
\end{align}
Equation~\eqref{eq:geo-u} integrates to
\begin{equation}
p_{u}=a^{2}\cosh^{2}\!\Bigl(\frac{v}{a}\Bigr)\dot u=\text{const}.
\label{eq:pu}
\end{equation}
Combining Eq.~\eqref{eq:pu} with the normalization
$2E=\cosh^{2}(v/a)\bigl(\dot v^{2}+a^{2}\dot u^{2}\bigr)$
yields
\begin{equation}
\dot v^{2}
=\frac{2E}{\cosh^{2}(v/a)}
-\frac{p_{u}^{2}}{a^{2}\cosh^{4}(v/a)}.
\label{eq:vdot}
\end{equation}
Introducing the dimensionless ratio
\begin{equation}
\Lambda=\frac{p_{u}}{a\sqrt{2E}},
\label{eq:LambdaDef}
\end{equation}
one obtains the first-order orbit equation
\begin{equation}
\frac{du}{dv}
=\pm\frac{\Lambda}{a \sqrt{\cosh\!\left(\frac{v}{a}\right)^2 - \Lambda^2}}.
\label{eq:orbit}
\end{equation}

The integral in Eq.~\eqref{eq:orbit} can be performed analytically, and the solutions are characterized by the parameter $\Lambda$. Detailed solutions are constructed in Appendix~\ref{appgeo}; here we collect the main results. Equation~\eqref{eq:orbit} admits four qualitatively distinct regimes depending on $\Lambda$:

\begin{itemize}
\item[{\bf (i)}] {\bf Meridional geodesics:} $\Lambda=0$.  
Here $du/dv=0$, so $u=\text{const}$ and the curve runs along a meridian, crossing the neck orthogonally.

\item[{\bf (ii)}] {\bf Trans--throat spirals (subcritical):} $0<|\Lambda|<1$.  
The square root in Eq.~\eqref{eq:orbit} is real for all $v$, and the geodesic passes smoothly through the neck:
\begin{equation}
u(v) - u_{0}
= \pm\,\Lambda\,
F\!\left(
\arcsin\!\left(
\frac{\sinh\!\left(\tfrac{v}{a}\right)}
{\sqrt{\cosh^{2}\!\left(\tfrac{v}{a}\right) - \Lambda^{2}}}
\right)
\;\middle|\;
\Lambda^{2}
\right),
\label{eq:u_of_v_Lambda_lt_1}
\end{equation}
where $F(\phi\,|\,m)$ is the elliptic integral of the first kind,
\begin{equation}
F(\phi\,|\,m) = \int_{0}^{\phi} \frac{d\theta}{\sqrt{1 - m\,\sin^{2}\theta}}.\nn
\end{equation}
For $|\Lambda|<1$, we have $\cosh^{2}\!\left(\tfrac{v}{a}\right) - \Lambda^{2} > 0$ for all $v$, and the expression is valid globally.

Alternatively, the solution can be expressed more compactly as
\begin{equation}
u(v)-u_{0}
=\pm \frac{i \Lambda \, F\left( \frac{i v}{a}, \frac{1}{1 - \Lambda^2} \right)}{\sqrt{1 - \Lambda^2}},
\label{eq:geo-L<1}
\end{equation}
which is real for $|\Lambda|<1$.

\item[{\bf (iii)}] {\bf Circular neck geodesics (critical):} $|\Lambda|=1$.  
In the limit $v\to 0$, one obtains the circular throat geodesic at $v=0$:
\[
v = 0, \quad u(\tau) = u_0 + \omega \tau, \quad |\Lambda| = 1.
\]

\item[{\bf (iv)}] {\bf Trapped one--sided geodesics (supercritical):} $|\Lambda|>1$.  
Here the motion is confined to $v\ge v_{\mathrm{tp}}$, where
$\cosh(v_{\mathrm{tp}}/a)=|\Lambda|$.
Integrating Eq.~\eqref{eq:orbit} gives
\begin{equation}
u(v) - u_{0}
= \pm
F\!\left(
\arcsin\!\left(
\frac{\sqrt{\cosh^{2}\!\left(\tfrac{v}{a}\right) - \Lambda^{2}}}
{\sinh\!\left(\tfrac{v}{a}\right)}
\right)
\;\middle|\;
\frac{1}{\Lambda^{2}}
\right),
\label{eq:geo-L>1}
\end{equation}
where again $F(\phi\,|\,m)$ denotes the elliptic integral of the first kind.
\end{itemize}

We now construct dipole trajectories using Eq.~\eqref{dyneq} specialized to $N=2$ with counter-rotating unit circulations. It is useful to express the geodesic parameter $\Lambda$ in Eq.~\eqref{eq:LambdaDef} in terms of the dipole initial data, in order to systematically generate trajectories in a given geodesic class. Following Drivas, Glukhovskiy, and Khesin~\cite{khesin2024}, we consider a dipole with strengths $\Gamma_1=+1$ and $\Gamma_2=-1$ and define the mean variables
\[
\bar{u} = \frac{u_1 + u_2}{2}, 
\qquad 
\bar{v} = \frac{v_1 + v_2}{2},
\]
with corresponding mean velocities
\[
\dot{\bar{u}} = \frac{\dot{u}_1 + \dot{u}_2}{2}, 
\qquad 
\dot{\bar{v}} = \frac{\dot{v}_1 + \dot{v}_2}{2}.
\]
The azimuthal momentum $p_u$ is then
\[
p_u = a^2 \cosh^2\!\left(\frac{\bar{v}}{a}\right) \dot{\bar{u}},
\]
and the energy $E$ is
\[
2E = \cosh^2\!\left(\frac{\bar{v}}{a}\right) \left(\dot{\bar{v}}^2 + a^2 \dot{\bar{u}}^2\right).
\]
The dimensionless ratio $\Lambda$ becomes
\[
\Lambda = \frac{p_u}{a \sqrt{2E}}
= \frac{a^2 \cosh^2\!\left(\frac{\bar{v}}{a}\right) \dot{\bar{u}}}{a \sqrt{\cosh^2\!\left(\frac{\bar{v}}{a}\right)\left(\dot{\bar{v}}^2 + a^2 \dot{\bar{u}}^2\right)}}.
\]

This expression characterizes the effective motion of the dipole in terms of the mean position $(\bar u,\bar v)$ and mean velocity $(\dot{\bar u},\dot{\bar v})$. In the tight-dipole regime $(\Delta u,\Delta v)\ll 1$, the limit must be taken in an ordered manner---first $\Delta u\to 0$ and subsequently $\Delta v\to 0$ reflecting the singular structure of the normalization in Eq.~\eqref{eq:LambdaDef}. Carrying out this procedure yields
\[
\Lambda
=
\cosh\!\left(\frac{\bar v}{a}\right)\,\mathrm{sgn}(\Delta v)
+ O(\text{dipole size}^2).
\]
A detailed derivation of this ordered limit, starting from the exact two-vortex equations and including all intermediate expansions, is provided in Appendix~\ref{app:LambdaFullDerivation}. More generally, away from the ordered meridional limit, the geodesic
parameter $\Lambda$ can be expressed directly in terms of the dipole
center velocity $(\dot{\bar u},\dot{\bar v})$ and the local metric
factor $h(\bar v)=\cosh(\bar v/a)$,
\[
\Lambda =
\frac{a\,h(\bar v)\,\dot{\bar u}}
{\sqrt{\dot{\bar v}^{\,2}+a^{2}\dot{\bar u}^{\,2}}}.
\]
This relation provides a practical way to initialize vortex dipoles so
that the dipole center follows a prescribed geodesic trajectory. From this expression, we can systematically engineer dipole trajectories that follow meridional geodesics, circular neck geodesics, trapped one-sided (supercritical) and trans throat subcritical geodesics. Figures~\ref{figgdsc1}--\ref{figgdsc3} summarize the numerical verification of the geodesic conjecture on the catenoid for three representative regimes of the dimensionless parameter $\Lambda$.

In Fig.~\ref{figgdsc1}, we consider a meridional trajectory corresponding to $\Lambda=0$, with initial conditions $(u_1,v_1,u_2,v_2) = (\epsilon,-2,-\epsilon,-2)$ and $\epsilon = 0.05$, for which the dipole center moves axially through the throat along $u=\text{const}$. The top panels show the time evolution of $(u(t),v(t))$ and the corresponding embedded trajectory on the catenoid, demonstrating agreement between the numerical vortex-pair evolution and the analytical meridional geodesic. In the $(u,v)$ trajectory plots, the green curve denotes the analytical geodesic solution, while the blue curve shows the corresponding vortex--dipole evolution. For clarity, the geodesic curve is intentionally terminated at an earlier value of the affine parameter so that both trajectories can be visualized distinctly in regions where they would otherwise overlap almost exactly. The lower panels plot the errors $\delta H(t)=|H(t)-H(0)|$ and $\delta J(t)=|J(t)-J(0)|$, which remain below $10^{-8}$ throughout the integration, confirming numerical conservation of both invariants.

Figure~\ref{figgdsc2} corresponds to the circular neck (throat) geodesic with $\Lambda=1$, obtained for initial conditions $(u_1,v_1,u_2,v_2)=(0,\epsilon,0,-\epsilon)$. Here the dipole remains localized near $v=0$, executing a closed orbit around the neck of the catenoid. The embedding panel shows the circular motion characteristic of the $|\Lambda|=1$ family. The conservation plots again show that both $H$ and $J$ remain constant to machine precision, demonstrating that the full vortex dynamics respect the Hamiltonian structure.

Finally, Fig.~\ref{figgdsc3} depicts a trapped one-sided geodesic with $|\Lambda|>1$, realized with initial data $(u_1,v_1,u_2,v_2) = (0,0.15+\epsilon,0,0.15-\epsilon)$ and $\epsilon=0.05$. In this regime, the trajectory remains confined to one side of the catenoid, as predicted by the analytic orbit equation~\eqref{eq:orbit}. The numerical path in the $(u,v)$ plane and its embedding on the catenoid match the theoretical turning point $v_{\mathrm{tp}} = a\cosh^{-1}|\Lambda|$, while $\delta H$ and $\delta J$ remain bounded below $10^{-7}$, confirming conservation of the Hamiltonian and azimuthal momentum.

Taken together, these three representative cases---$\Lambda=0$, $\Lambda=1$, and $|\Lambda|>1$---provide a direct numerical demonstration that vortex--dipole trajectories follow the meridional, circular, and trapped (supercritical) geodesics of the catenoid metric, thereby validating the correspondence and conservation properties of the Hamiltonian system.
\section{Demonstration of Direct and Exchange Scattering of Dipoles}
\label{scattering}
\begin{figure}[htbp!]
\begin{tabular}{lcccccccc}
\includegraphics[height=0.25\textheight]{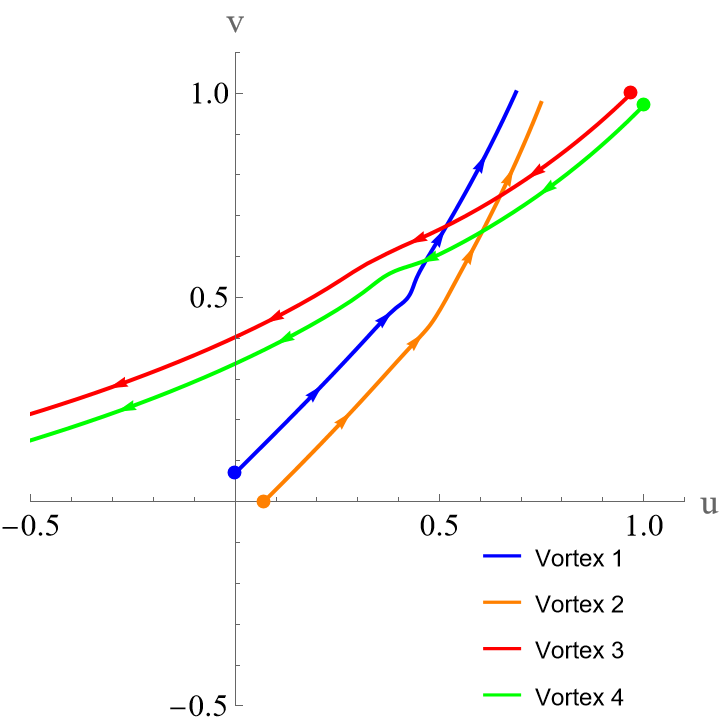}&&
\includegraphics[height=0.25\textheight]{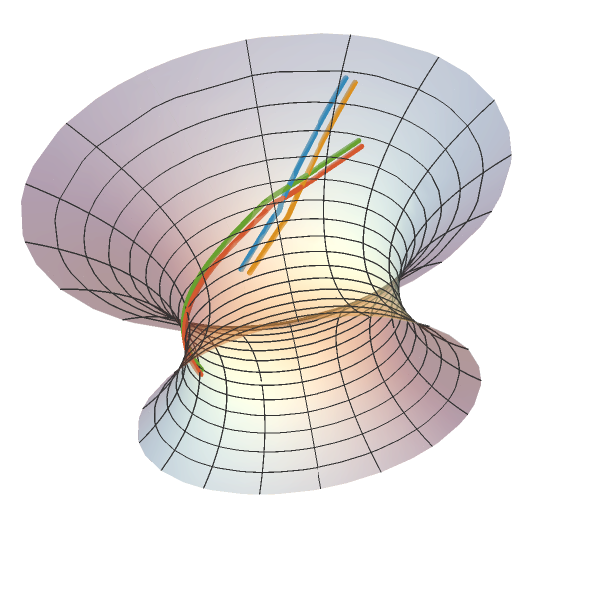}\\
\includegraphics[height=0.18\textheight]{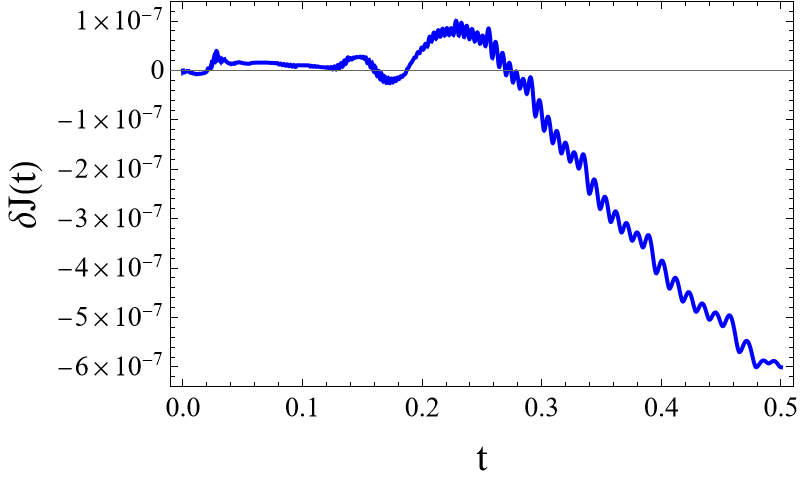}&&
\includegraphics[height=0.18\textheight]{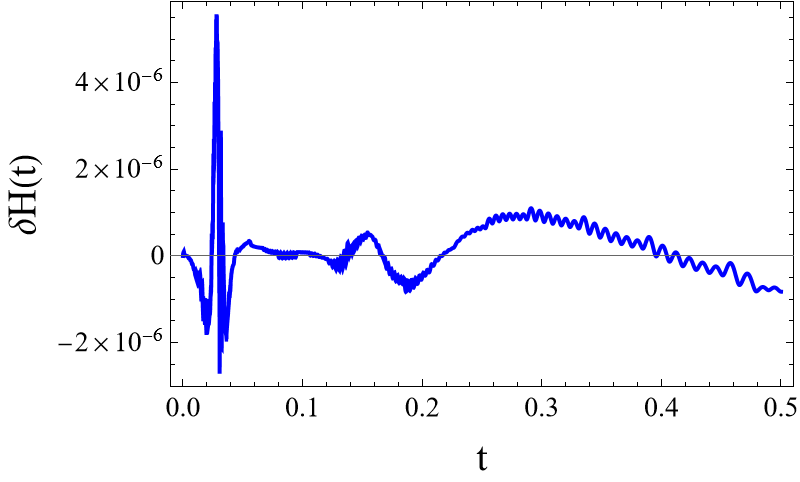}
\end{tabular}
\caption{
Direct scattering of two dipoles on a catenoid surface. 
The dipoles are initialized symmetrically about the diagonal with a small offset, with the first dipole located near the origin  in the $(u,v)$ coordinate plane and the second approaching from the vicinity of $(1,1)$. 
A small parameter $\epsilon = 0.07, \delta = 0.03$ defines the initial configuration 
$\{(0,\epsilon),\,(\epsilon,0),\,(1-\delta,1),\,(1,1-\delta)\}$, with vortex strengths $\Gamma=\{-1,1,1,-1\}$. 
The top panels show the trajectories at the final integration time $t_f = 0.5$, both in the $(u,v)$ coordinate plane and mapped onto the catenoid surface. 
The lower panels display the temporal evolution of the conserved quantities $J$ and $H$, demonstrating their numerical conservation throughout the simulation.
}
\label{figgdsc4} 
\end{figure}

\begin{figure}[htbp!]
\begin{tabular}{lcccccccc}
\includegraphics[height=0.25\textheight]{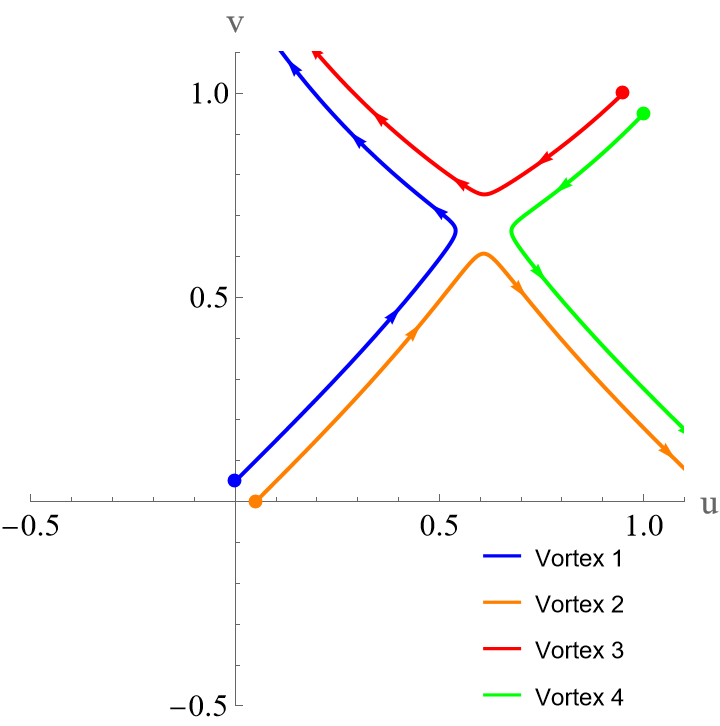}&&
\includegraphics[height=0.25\textheight]{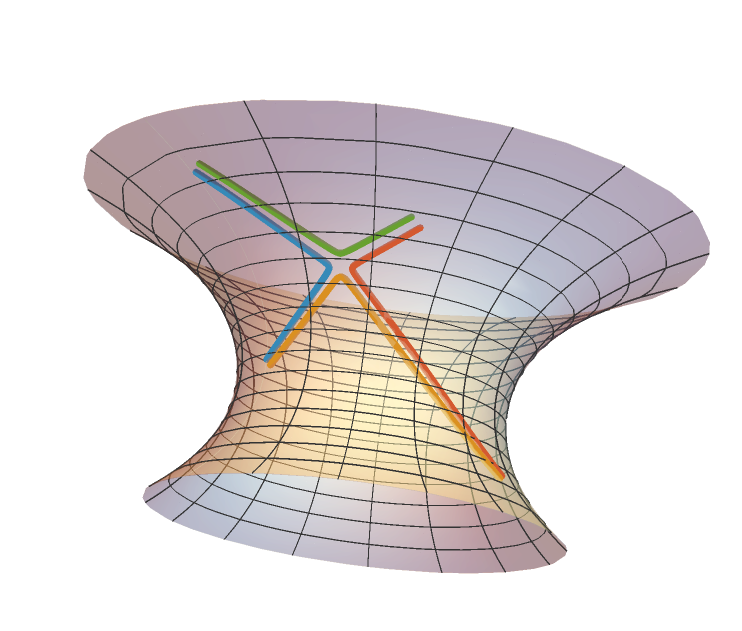}\\
\includegraphics[height=0.15\textheight]{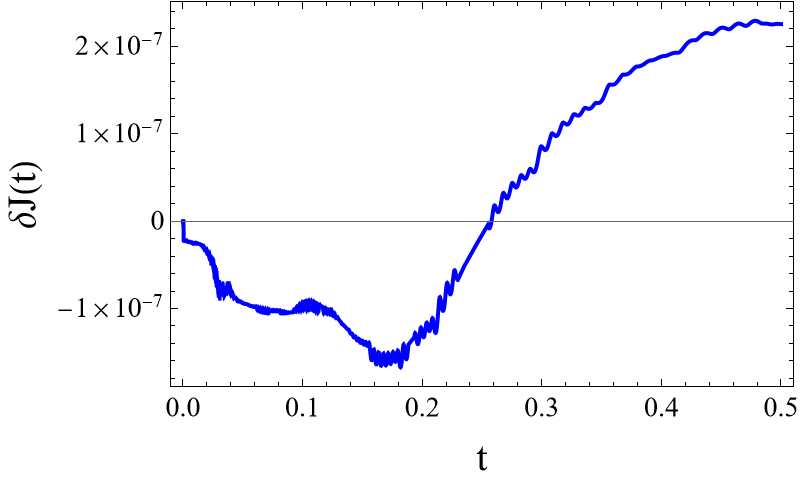}&&
\includegraphics[height=0.15\textheight]{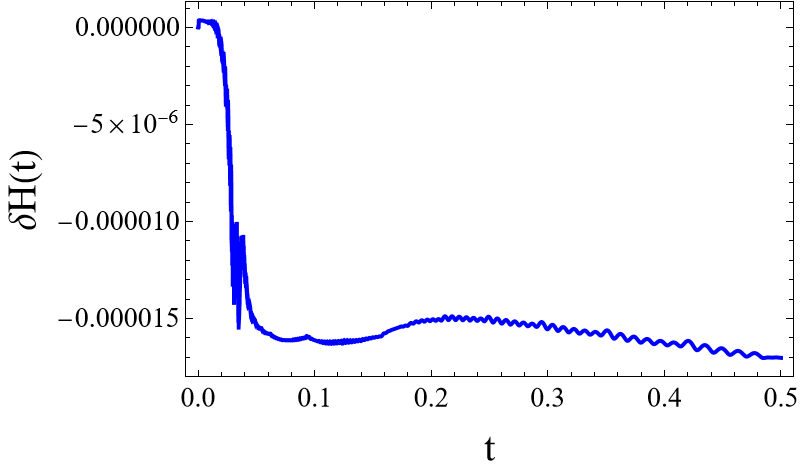}\\
\end{tabular}
 \caption{Exchange scattering of two dipoles on a catenoid surface. 
The dipoles are initialized symmetrically about the diagonal, with the first dipole located near the origin  in the $(u,v)$ coordinate plane and the second approaching from the vicinity of $(1,1)$. 
A small parameter $\epsilon = \delta = 0.05$ defines the initial configuration 
$\{(0,\epsilon),\,(\epsilon,0),\,(1-\delta,1),\,(1,1-\delta)\}$, with vortex strengths $\Gamma=\{-1,1,1,-1\}$. 
The top panels show the trajectories at the final integration time $t_f = 0.5$, both in the $(u,v)$ coordinate plane and mapped onto the catenoid surface. 
The lower panels display the temporal evolution of the conserved quantities $J$ and $H$, demonstrating their numerical conservation throughout the simulation.}

 \label{figgdsc5} 
\end{figure}

We now demonstrate that the  dipole dynamics described by Eq.~(\ref{dyneq}) on the catenoid admit the two fundamental types of two–body scattering processes: the \emph{direct} and \emph{exchange} scattering of vortex dipoles.  Study of such collisions is interesting  in the context of BEC formation, see Ref.~\cite{Neely2010,Freilich2010} in curved trap geometries. In contrast to the flat case, here the Gaussian curvature of the surface modifies both the impact and the post-collision trajectories, providing a geometric control parameter for the scattering.

\paragraph{Direct scattering.}
Figure~\ref{figgdsc4} shows the direct–scattering event for two oppositely oriented dipoles initialized near the catenoid throat. The initial configuration is described by the coordinates $\{(0,\epsilon),(\epsilon,0),(1-\delta,1),(1,1-\delta)\}$ with strengths $\Gamma=\{-1,1,1,-1\}$.
The small parameters are chosen to introduce a small off–axis asymmetry \(\epsilon = 0.07\), \(\delta = 0.03\), and the final integration time \(t_f = 0.5\). 
The left figure depict the vortex trajectories  in the $(u,v)$–plane and the right figure shows the scattering on the catenoid surface. As the two dipoles approach one another from opposite sides, they interact briefly and then separate, each preserving its internal identity and orientation. 
The $(u,v)$ projection displays the characteristic right–angle scattering familiar from planar hydrodynamics, now modulated by the non-uniform metric factor \(h(v) = \cosh(v/a)\). 
Throughout the encounter the Hamiltonian and azimuthal momentum remain conserved to within a relative error of \(10^{-7}\), confirming that the event is an exact Hamiltonian flow on the catenoid. 
\paragraph{Exchange scattering.}
Figure~\ref{figgdsc5} depicts the exchange–scattering regime obtained  in the initial  locations of the dipoles
by removing the small off-axis symmetry of the direct scattering channel ie. \(\epsilon = \delta= 0.05\) and \(t_f = 0.5\). 
Here the dipoles collide obliquely and exchange partners: after the encounter, each vortex becomes bound to the opposite sign vortex from the other dipole, forming two new dipoles that propagate away from the throat region. 
The embedded–surface trajectories reveal a distinct crossing and recombination pattern of the vortex lines. 
Once again, both $H$ and $J$ remain conserved to numerical precision, demonstrating that the exchange process, although topologically nontrivial, still corresponds to a symplectic dynamics in the four–vortex phase space. \\
Taken together, Figs.~\ref{figgdsc4} and~\ref{figgdsc5} establish that the catenoid supports the full range of classical two–dipole scattering behaviours. 
In both cases, conservation of the Hamiltonian and azimuthal momentum to machine accuracy demonstrates the exact symplectic nature of the evolution governed by Eq.~(\ref{dyneq}). 
These results extend the canonical planar picture of vortex–dipole collisions to curved manifolds, showing explicitly how the geometry of the catenoid—through its negative Gaussian curvature—acts as a tunable control of the scattering angles. To further distinguish the direct and exchange scattering channels, we can define the dipole orientation $\alpha$ as the angle of the dipole axis (from negative to positive vortex) in the local orthonormal frame $\{\hat e_u,\hat e_v\}$ of the catenoid metric, and the impact parameter as the transverse projection (with respect to the metric inner product) of the initial centre--centre separation onto the direction perpendicular to the initial relative self-propulsion velocity.  For the representative initial conditions shown in Figs.~\ref{figgdsc4} and \ref{figgdsc5}, both configurations correspond to head-on collisions (vanishing impact parameter to numerical precision), and their relative orientations differ only at the $10^{-3}$ level. 
The distinction between direct and exchange scattering therefore does not arise from geometric differences in impact parameter or orientation, but from the fact that the two initial conditions lie on different invariant manifolds of the conserved azimuthal momentum $J$. For the present initial data $\{(0,\epsilon),(\epsilon,0),(1-\delta,1),(1,1-\delta)\}$ 
with $\Gamma=\{-1,1,1,-1\}$, the conserved azimuthal momentum takes the explicit form
\[
J(\epsilon,\delta)
=
\frac{a}{4}
\left[
2(\delta-\epsilon)
+
a\,\sinh\!\left(\frac{2}{a}\right)
-
a\left(
\sinh\!\left(\frac{2-2\delta}{a}\right)
+
\sinh\!\left(\frac{2\epsilon}{a}\right)
\right)
\right].
\]
so that the two configurations in Figs.~\ref{figgdsc4} and \ref{figgdsc5} lie on distinct $J$-manifolds despite their nearly identical geometric impact parameters and orientations. A detailed study of all scattering channels on the catenoid characterized by $J$ is left for a more detailed future communication. 
\begin{table}[t]
\centering
\small
\caption{Primary modes of vortex--dipole motion on the catenoid studied in this work and the controlling dimensionless parameters.
Here $a$ is the throat radius, $\ell$ the (geodesic) dipole size, $h(v)=\cosh(v/a)$, and
$\Lambda=p_u/(a\sqrt{2E})$ labels the geodesic family in the dipole limit.}
\label{tab:modes_catenoid}
\begin{tabular}{p{0.22\linewidth} p{0.26\linewidth} p{0.34\linewidth} p{0.14\linewidth}}
\hline
\textbf{Mode} & \textbf{Dynamical level} & \textbf{Key dimensionless controls} & \textbf{Outcome} \\
\hline

Meridional geodesic
& Dipole limit (tight vortex pair)  (Sec.~\ref{geodesics})
& $\Lambda=0$; centroid height $\bar v_0/a$; dipole tightness $\varepsilon\sim(\Delta u,\Delta v)\ll 1$
& Crosses throat, $u=\mathrm{const}$ \\[2pt]

Neck circle (critical)  
& Dipole limit (tight vortex pair) (Sec.~\ref{geodesics})
& $|\Lambda|=1$; $\varepsilon\ll 1$
& Localized near $v=0$ (throat orbit) \\[2pt]
Subcritical  geodesic
& Dipole limit (tight vortex pair)  (Sec.~\ref{geodesics})
& $0<|\Lambda|<1$; centroid height $\bar v_0/a$; $\varepsilon\ll1$
& Trans--throat  trajectory \\[2pt]

One--sided trapped (supercritical) 
& Dipole limit (tight vortex pair) (Sec.~\ref{geodesics})
& $|\Lambda|>1$; turning point $\cosh(v_{\rm tp}/a)=|\Lambda|$; $\varepsilon\ll 1$
& Confined to one lobe (no throat crossing) \\[4pt]

Direct dipole--dipole scattering 
& Full Hamiltonian $4$--vortex dynamics  (Sec.~\ref{scattering})
& distinct $J$ and $H$ class
& Dipoles separate with identity preserved \\[2pt]

Exchange scattering
& Full Hamiltonian $4$--vortex dynamics (near field) (Sec.~\ref{scattering})
& distinct $J$ and $H$ class
& Partner swap; outgoing dipoles re-paired \\[4pt]

Finite-sized coherent dipole motion
& Reduced $N$--dipole model (Sec.~\ref{dipolesystem})
& dipole size $\ell/a\ll 1$; separation $R/\ell\gg 1$ (far field); local curvature via $\sech(v/a),\tanh(v/a)$
& Curvature--modulated self-propulsion with far-field interactions and parallel-transport precession \\

\hline
\end{tabular}
\end{table}
\section{Demonstration of Collective Rotation for Co-Rotating Pairs}
\label{cluster}

\begin{figure}[htbp!]
\begin{tabular}{lcccccccc}
\includegraphics[height=0.25\textheight]{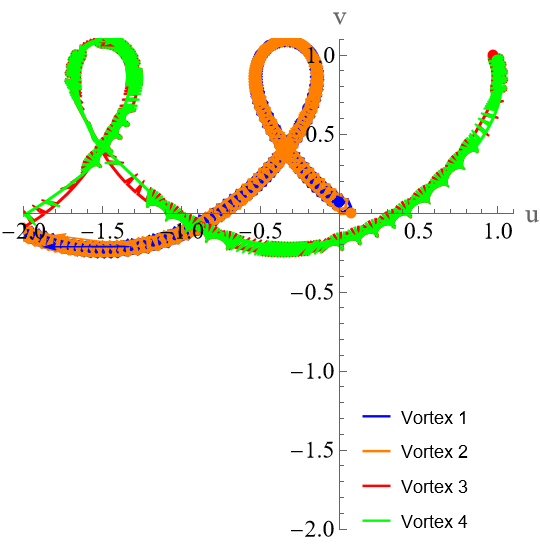}&&
\includegraphics[height=0.25\textheight]{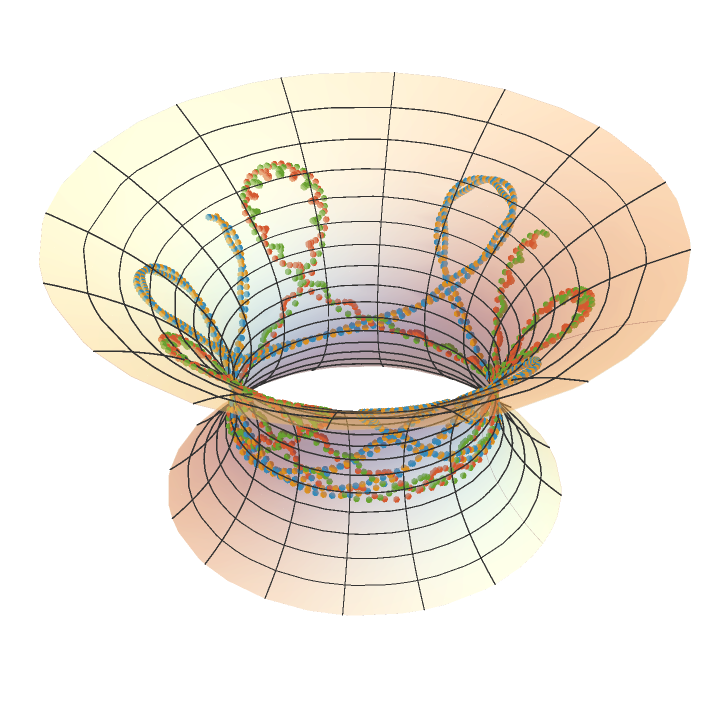}
\end{tabular}
\caption{
Co-rotating four-vortex dynamics on a catenoid surface. 
The vortices are initialized in the direct-scattering configuration 
$\{(0,\epsilon),\, (\epsilon,0),\, (1-\delta,1),\, (1,1-\delta)\}$ 
with identical circulations $\Gamma_i = 1$, $\epsilon = 0.07$, $\delta = 0.03$, and $t_f = 10$. 
Unlike the dipole configurations, the uniform circulation leads to collective rotational motion (with azimuthal drift) and intricate looping trajectories, shown in the Cartesian grid (left) and on the catenoid surface (right).
}
\label{samesign_ex} 
\end{figure}

\begin{figure}[htbp!]
\begin{tabular}{lcccccccc}
\includegraphics[height=0.25\textheight]{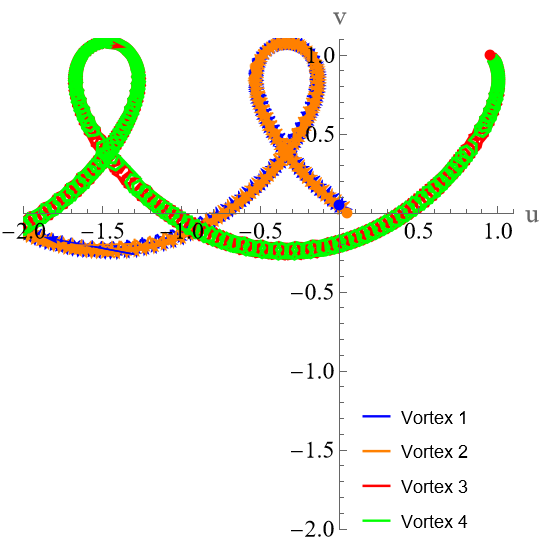}&&
\includegraphics[height=0.25\textheight]{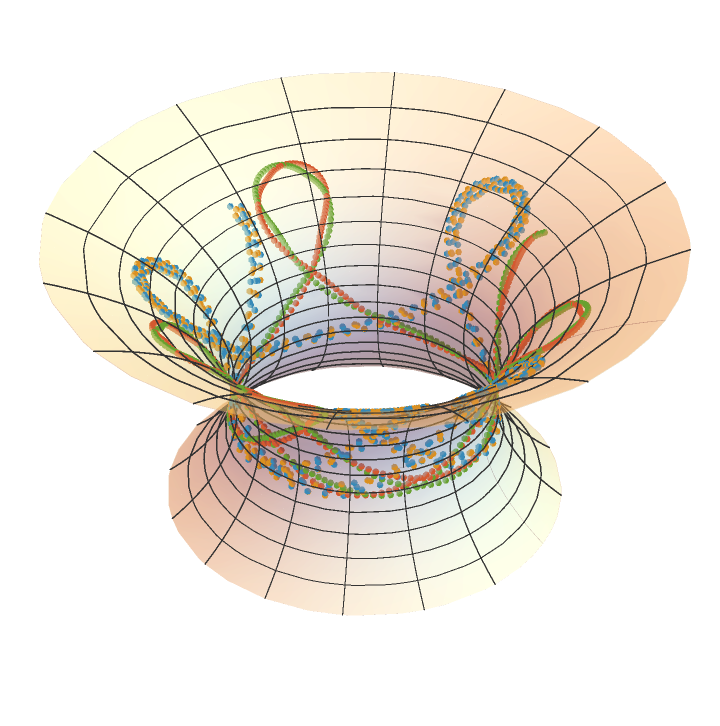}
\end{tabular}
\caption{
Co-rotating four-vortex dynamics on a catenoid surface. 
The vortices are initialized in the exchange-scattering configuration 
$\{(0,\epsilon),\, (\epsilon,0),\, (1-\delta,1),\, (1,1-\delta)\}$ 
with identical circulations $\Gamma_i = 1$, $\epsilon = 0.05$, $\delta = 0.05$, and $t_f = 10$. 
Unlike the dipole configurations, the uniform circulation leads to collective rotational motion (with azimuthal drift) and intricate looping trajectories, shown in the Cartesian grid (left) and on the catenoid surface (right).
}
\label{samesign_dsc} 
\end{figure}

In contrast to dipoles, where opposite circulations generate translational self-propulsion, co-rotating pairs with identical circulations produce mutual rotation about their midpoint, resulting in a bound collectively rotating state. We now explore the same geometric configurations considered for dipoles in the previous section, but with co-rotating vortices.

A comprehensive analysis of co-rotating vortex clusters on surfaces of variable negative curvature, including full Hamiltonian reduction, stability theory, and nonlinear phase portraits, will be presented in a forthcoming dedicated work~\cite{samanta2026}. Here, we summarize only the key structural features relevant for comparison with the dipole dynamics.

For equal circulations $\Gamma_1=\Gamma_2=\Gamma>0$, the system admits a symmetric diametric co-rotating state
\[
\Delta u=\pi, \qquad \Delta v=0, \qquad v_1=v_2=V_0,
\]
which constitutes a relative equilibrium. The vortices remain at fixed meridional height and rotate rigidly with angular velocity
\[
\Omega(V_0)
=
\frac{\Gamma}{4\pi a^2}
\frac{\tanh(V_0/a)}{\cosh^2(V_0/a)}
=
\frac{\Gamma}{16\pi}
\frac{K'(V_0)}{\sqrt{-K(V_0)}},
\]
so that the curvature scale $a^{-2}$ sets the natural frequency. 
Here $K(V)$ denotes the Gaussian curvature of the catenoid,
\[
K(V)=-\frac{1}{a^2\cosh^4(V/a)},
\]
and $K'(V)=dK/dV$. In this symmetric state, there is no meridional drift.

Linearization shows that the configuration is hyperbolic, with instability entirely curvature-driven and vanishing in the planar limit. However, because the reduced system has one degree of freedom and conserves both energy and angular momentum, the nonlinear dynamics remains globally bounded: perturbations grow exponentially along the unstable manifold but remain confined to invariant energy contours.

Away from the perfectly symmetric state, the reduced dynamics in $(\Delta v,\Delta u)$ has one degree of freedom and is generically periodic for bounded energies. The collective azimuthal motion arises via Hamiltonian reconstruction,
\[
\dot U(t)=\Omega_0+\delta\Omega(t),
\]
with nonzero mean $\Omega_0=\langle\dot U\rangle$ and bounded periodic modulation $\delta\Omega(t)$. Consequently,
\[
U(t)=\Omega_0 t + U_{\rm osc}(t),
\]
describing collective rotation with oscillatory separation and a curvature-induced meridional adjustment governed by conservation of angular momentum $J$. The drift vanishes in the planar limit $a\to\infty$, confirming its geometric origin.

We may also measure the vortex–vortex separation using the Euclidean chord distance in the standard $\mathbb{R}^3$ embedding of the catenoid,
\[
d_{\mathbb R^3}(t)=\|X(u_1(t),v_1(t)) - X(u_2(t),v_2(t))\|,
\]
which exhibits oscillatory behavior reflecting the combined azimuthal rotation and meridional motion of the vortices through modulation of the embedding radius $a\cosh(v/a)$.

We now demonstrate the collective rotational dynamics, together with the accompanying azimuthal drift, by initializing four co-rotating vortices in the same geometric configuration as the dipoles shown in Fig.~\ref{figgdsc4}. Figures~\ref{samesign_ex} and~\ref{samesign_dsc} present representative examples of this co-rotating regime.

Figure~\ref{samesign_ex} shows the evolution of symmetric pairs of vortices with identical circulation $\Gamma=+1$, initially placed in the same configuration as the dipoles in Fig.~\ref{figgdsc4}, with integration time $t_f = 10$. The panels display the vortex trajectories in the $(u,v)$ plane and on the embedded catenoid surface. The vortices execute uniform counter-rotating motion about the neck, maintaining nearly constant separation and forming steady co-rotating pairs with azimuthal drift.

The embedding view highlights the circular orbits wrapping around the throat, while the $(u,v)$ plot exhibits oscillations in height $v$, induced by the metric factor $h(v)=\cosh(v/a)$. Figures~\ref{samesign_ex} and~\ref{samesign_dsc} together demonstrate the existence of stable, collectively rotating vortex pairs on negatively curved surfaces. Unlike the dipole motion discussed in Sec.~\ref{scattering}, where opposite circulations lead to translation and scattering, same-sign pairs undergo rigid rotation about the surface neck, driven by the intrinsic curvature of the catenoid.

\section{Construction of the finite--dipole dynamical system and self--propulsion terms}
\label{dipolesystem}
\begin{figure}[htbp!]
    \centering
    \includegraphics[width=0.75\textwidth]{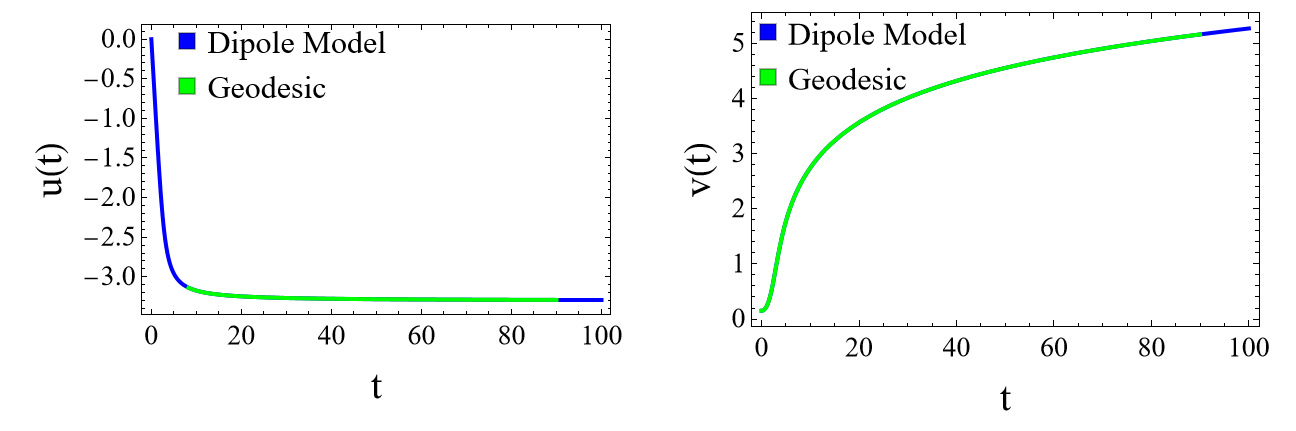}\\[1em]
    (a) Comparison of $u(t)$ and $v(t)$ between dipole model and geodesic motion\\[0.5em]
    \includegraphics[width=0.75\textwidth]{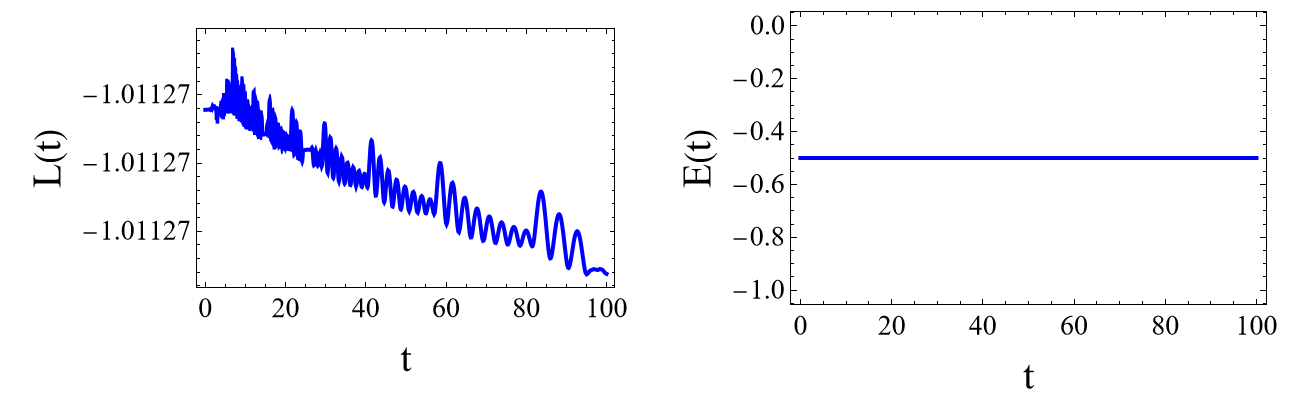}\\[1em]
    (b) Conserved quantities $L(t)$ and $E(t)$ along the geodesic\\[0.5em]
    \caption{
        Comparison between the finite--dipole model and the corresponding geodesic motion on a catenoid. 
        (\textbf{Top}) Evolution of the dipole coordinates $u(t)$ and $v(t)$ obtained from the self--propelled dipole model (blue) and the analytical geodesic motion (green). 
        (\textbf{Bottom}) The angular momentum $L(t)$ and energy $E(t)$ remain constant, confirming that the reduced dipole model preserves the expected invariants of geodesic motion.
    }
    \label{fig:dpsystem_geodesic}
\end{figure}
In Sec.~\ref{ham} we derived the exact Hamiltonian dynamics of $N$ point vortices on the catenoid. 
In Sec.~\ref{scattering} we used this formulation to analyze full vortex interactions, including near-field interaction events such as exchange scattering. 
We now construct a reduced dynamical model for $N$ \emph{finite-sized vortex dipoles}, obtained from a systematic small--separation expansion of the $2N$ exact vortex equations. The reduced description assumes that each dipole remains a coherent object of small but fixed geodesic size $\ell_n$ (imposed as  a constraint), with $\ell_n$ small compared to the local radius of curvature and to the inter-dipole separation. In this regime the dominant effects are (i) curvature--corrected self--propulsion, (ii) far-field dipole--dipole interactions, and (iii) geometric rotation induced by parallel transport. Near-field interaction events involving partner exchange lie outside the regime of validity of the finite--dipole approximation and require the full vortex Hamiltonian dynamics discussed in Sec.~\ref{scattering}. 
We therefore restrict attention here to the asymptotic regime in which each dipole remains a coherent object of small geodesic size. The resulting reduced system provides a simple but dynamically consistent model of interacting finite--sized dipoles on the catenoid. 
Such a formulation is also relevant for weakly interacting vortex dipoles in Bose--Einstein condensates confined to curved trap geometries, see \cite{Tchieu2012} for similar investigations in planar geometries.  Starting from the catenoid metric and its associated orthonormal frame, we now derive the curvature--corrected self--propulsion terms together with the coupled evolution equations for the dipole centre and its orientation. 
The catenoid metric is
\begin{equation}
ds^2 = a^2 h(v)^2\,du^2 + h(v)^2\,dv^2,
\qquad 
h(v) = \cosh\!\left(\tfrac{v}{a}\right),
\end{equation}
we first construct the orthonormal basis vectors 
\begin{equation}
\hat e_u = \frac{1}{a h}\,\partial_u,
\qquad
\hat e_v = \frac{1}{h}\,\partial_v.
\end{equation}
The Levi--Civita connection one--form in this orthonormal frame is
\begin{equation}
\omega(\partial_u) = \tanh(v/a), 
\qquad 
\omega(\partial_v)=0,
\end{equation}
so that a tangent vector parallel transported along a trajectory with velocity components $(\dot u,\dot v)$ rotates at the rate
\begin{equation}
\omega(X) = \tanh(v/a)\,\dot u.
\end{equation}
For convenience, let us recall that a system of  $2N$ point vortices with circulations $\Gamma_i$, the equations of motion in $(u,v)$ coordinates are
\begin{subequations}
\begin{align}
\dot v_i &=
\frac{1}{4\pi a\,h_i^2}
\sum_{j\neq i}\Gamma_j
\frac{\sin(u_i - u_j)}{F_{ij}},\nn\\[4pt]
\dot u_i &=
-\frac{1}{4\pi a^2\,h_i^2}
\sum_{j\neq i}\Gamma_j
\frac{\sinh\!\big(\tfrac{v_i - v_j}{a}\big)}{F_{ij}}
+\frac{1}{4\pi a^2\,h_i^2}\,\Gamma_i\tanh\!\big(\tfrac{v_i}{a}\big),\nn\\
\label{eq:pv-u}
\end{align}
\end{subequations}
where
\begin{equation}
F_{ij} = 
\cosh\!\Big(\frac{v_i-v_j}{a}\Big) - \cos(u_i - u_j),
\qquad
h_i = \cosh(v_i/a).\nn
\end{equation}
Each dipole $n$ consists of two vortices of equal and opposite circulation,
\(\Gamma_{n,+}=+1\) and \(\Gamma_{n,-}=-1\).
The two vortices are separated by a small geodesic distance $\ell_n$ along the tangent
\[
\hat{\mathbf t}_n = \cos\alpha_n\,\hat e_u + \sin\alpha_n\,\hat e_v,
\]
where $\alpha_n$ is the orientation angle measured from the $u$--direction.
To first order in $\ell_n$, the coordinates of the $\pm$ vortices are
\begin{equation}
u_{n,\pm} = u_n \pm \frac{\ell_n}{2 a h_n}\cos\alpha_n,
\qquad
v_{n,\pm} = v_n \pm \frac{\ell_n}{2 h_n}\sin\alpha_n,
\qquad
h_n = h(v_n).
\label{eq:uvpm}
\end{equation}
\paragraph{Self-Propulsion}
The self--propelled motion of a dipole arises from the mutual advection of its two constituent vortices. 
The velocity of the $+$ vortex due to its $-$ partner is
\begin{subequations}
\begin{align}
\dot u_{n,+}^{(\text{partner})}
&= -\frac{\Gamma_{n,-}}{4\pi a^2\,h(v_{n,+})^2}
\frac{\sinh\!\big(\tfrac{v_{n,+}-v_{n,-}}{a}\big)}{F_{(n,+)(n,-)}} +\frac{1}{4\pi a^2\,h(v_{n,+})^2}\,\Gamma_{n,+}\tanh\!\big(\tfrac{v_{n,+}}{a}\big),\\[3pt]
\dot v_{n,+}^{(\text{partner})}
&= \frac{\Gamma_{n,-}}{4\pi a\,h(v_{n,+})^2}
\frac{\sin(u_{n,+}-u_{n,-})}{F_{(n,+)(n,-)}},
\end{align}
\end{subequations}
and similarly with $+$ replaced by $-$. Averaging these two velocities gives the self--propulsion velocity of the dipole centre:
\begin{align}
\dot{u}_{n}^{(\mathrm{self})}
&= \frac{1}{8\pi a^{2}}
\bigg[
\frac{\sinh\!\big(\tfrac{v_{n,+}-v_{n,-}}{a}\big)}
     {h(v_{n,+})^{2}\,F_{(n,+)(n,-)}}
-
\frac{\sinh\!\big(\tfrac{v_{n,-}-v_{n,+}}{a}\big)}
     {h(v_{n,-})^{2}\,F_{(n,-)(n,+)}}
+
\frac{\tanh\!\big(\tfrac{v_{n,+}}{a}\big)}{h(v_{n,+})^{2}}
-
\frac{\tanh\!\big(\tfrac{v_{n,-}}{a}\big)}{h(v_{n,-})^{2}}
\bigg],\nn\\[6pt]
\dot{v}_{n}^{(\mathrm{self})}
&= -\,\frac{1}{8\pi a}
\bigg[
\frac{\sin\!\big(u_{n,+}-u_{n,-}\big)}
     {h(v_{n,+})^{2}\,F_{(n,+)(n,-)}}
+
\frac{\sin\!\big(u_{n,-}-u_{n,+}\big)}
     {h(v_{n,-})^{2}\,F_{(n,-)(n,+)}}
\bigg].
\label{eq:self-propulsion}
\end{align}
Using the displacements from Eq.~\eqref{eq:uvpm},
\[
\Delta u_n = u_{n,+}-u_{n,-} = \frac{\ell_n}{a h_n}\cos\alpha_n,
\qquad
\Delta v_n = v_{n,+}-v_{n,-} = \frac{\ell_n}{h_n}\sin\alpha_n,
\]
we obtain to leading order in $\ell_n$, the self--propulsion velocity components are (obtained in \textit{Mathematica})
\begin{equation}
\label{selfp_truncated}
\begin{aligned}
\dot u_i^{(\text{self})}
&=\frac{\mu_i\,\sech\!\big(\tfrac{v_i}{a}\big)\sin\alpha_i}{2a\pi\,\ell_i}
+ O(\ell_i),\\[6pt]
\dot v_i^{(\text{self})}
&= -\frac{\mu_i\cos\alpha_i\,\sech\!\big(\tfrac{v_i}{a}\big)}{2\pi\,\ell_i}
+ O(\ell_i),
\end{aligned}
\end{equation}
where $\mu_i$ represent the strength of dipole $i$. Note that the self-propulsion described by the above equation is orthogonal to the dipole axis, as expected. The full expressions are carried out in \textit{Mathematica} and presented in Appendix Sec~(\ref{fullexp}). 
\paragraph{External advection}
Using the compact notation
\[
F_{(n,\pm)(m,r)} = 
\cosh\!\Big(\frac{v_{n,\pm}-v_{m,r}}{a}\Big)
- \cos(u_{n,\pm}-u_{m,r}).
\]
the velocity of the + vortex of dipole $n$ induced by all other dipoles $m\neq n$ is

\begin{subequations}
\begin{align}
\dot u_{n,+}^{(\text{ext})}
&= -\frac{1}{4\pi a^2\,h_n^2}
\sum_{m\neq n}\sum_{r=\pm}
\Gamma_{m,r}
\left[
\frac{\sinh\!\big(\tfrac{v_{n,+}-v_{m,r}}{a}\big)}{F_{(n,+)(m,r)}}
\right],\nn\\[4pt]
\dot v_{n,+}^{(\text{ext})}
&= \frac{1}{4\pi a\,h_n^2}
\sum_{m\neq n}\sum_{r=\pm}
\Gamma_{m,r}
\left[
\frac{\sin(u_{n,+}-u_{m,r})}{F_{(n,+)(m,r)}}
\right]\nn
\end{align}
\end{subequations}
and similarly for the negative vortex. Note that since we have already incorporated the vortex self-interaction term in the self-propulsion, we exclude this contribution in the external advection. Averaging over the two  vortices in the n-th dipole, we have the total external contribution to the mean advection:
\begin{subequations}
\label{eq:external}
\begin{align}
\dot v_n^{(\text{ext})}
&= \frac{1}{8\pi a\,h_n^2}
\sum_{m\neq n}\sum_{r=\pm}
\Gamma_{m,r}
\left[
\frac{\sin(u_{n,+}-u_{m,r})}{F_{(n,+)(m,r)}}
+\frac{\sin(u_{n,-}-u_{m,r})}{F_{(n,-)(m,r)}}
\right],\\[4pt]
\dot u_n^{(\text{ext})}
&= -\frac{1}{8\pi a^2\,h_n^2}
\sum_{m\neq n}\sum_{r=\pm}
\Gamma_{m,r}
\left[
\frac{\sinh\!\big(\tfrac{v_{n,+}-v_{m,r}}{a}\big)}{F_{(n,+)(m,r)}}
+\frac{\sinh\!\big(\tfrac{v_{n,-}-v_{m,r}}{a}\big)}{F_{(n,-)(m,r)}}
\right],
\end{align}
\end{subequations}
with
\[
F_{(n,\pm)(m,r)} = 
\cosh\!\Big(\frac{v_{n,\pm}-v_{m,r}}{a}\Big)
- \cos(u_{n,\pm}-u_{m,r}).
\]
\paragraph{Evolution of the orientation angle.}
The orientation $\alpha_n$ evolves due to (i) the differential velocity of the two vortices (local shear), (ii) external torques from other dipoles and (iii) the geometric parallel transport on the curved surface, see more details on orientation dynamics in Appendix Sec.~(\ref{app:orientation}).
The first contribution is given by 
\begin{equation}
\dot{\alpha}_n^{(\text{self})}
=
\frac{\mu_n}{\ell_n}
\big(
 -\sin\alpha_n\,A_n
 + \cos\alpha_n\,B_n
\big),\nn
\end{equation}
\begin{align}
A_n &= a\!\left[
\cosh\!\left(\tfrac{v_{n,+}}{a}\right) u_{n,+}^{(\text{partner})}
- \cosh\!\left(\tfrac{v_{n,-}}{a}\right) u_{n,-}^{(\text{partner})}
\right],\nn \\[4pt]
B_n &= 
\cosh\!\left(\tfrac{v_{n,+}}{a}\right) v_{n,+}^{(\text{partner})}
- \cosh\!\left(\tfrac{v_{n,-}}{a}\right) v_{n,-}^{(\text{partner})}.\nn
\end{align}
The second contribution due to the external torques from the other dipoles is given by
\begin{equation}
\dot{\alpha}_n^{(\text{ext})}
=
\frac{\mu_n}{\ell_n}
\big(
 -\sin\alpha_n\,A_n^{(\text{ext})}
 + \cos\alpha_n\,B_n^{(\text{ext})}
\big),\nn
\end{equation}
\begin{align}
A_n^{(\text{ext})} &= a\!\left[
\cosh\!\left(\tfrac{v_{n,+}}{a}\right) u_{n,+}^{(\text{ext})}
- \cosh\!\left(\tfrac{v_{n,-}}{a}\right) u_{n,-}^{(\text{ext})}
\right],\nn \\[4pt]
B_n^{(\text{ext})} &= 
\cosh\!\left(\tfrac{v_{n,+}}{a}\right) v_{n,+}^{(\text{ext})}
- \cosh\!\left(\tfrac{v_{n,-}}{a}\right) v_{n,-}^{(\text{ext})}.\nn
\end{align}
Along with this, we also need to add the contribution to rotational contribution due to parallel transport,  more details on this appears in Appendix Sec.~(\ref{app:orientation}). The complete rotation rate is given by the sum of all three contributions:
\begin{equation}
\dot\alpha_n 
 = \dot\alpha_n^{(\text{self})}
+ \dot\alpha_n^{(\text{ext})}
+\tanh\!\big(\tfrac{v_n}{a}\big)\dot u_n.
\label{eq:alpha}
\end{equation}
The self term vanishes at leading order since the two partner vortices generate equal and opposite velocities, producing translation without rotation. 
The subleading correction is (obtained in \textit{Mathematica})
\begin{equation}
\label{eq:third_term_raw_va}
\dot\alpha_i^{(\text{self})}\;=\; -\frac{\mu_i\,\sech^3\!\big(\tfrac{v_i}{a}\big)\,\sin\alpha_i\,
\tanh\!\big(\tfrac{v_i}{a}\big)\,\ell_i}{24\,a^{3}\pi}\,
\Big(\cos^2\alpha_i - 6\sin^2\alpha_i + 6\sin^2\alpha_i\,\tanh^2\!\big(\tfrac{v_i}{a}\big)\Big)
+ O(\ell_i^2).\nn
\end{equation}
The full expressions are carried out in \textit{Mathematica} and presented in Appendix Sec~(\ref{fullexp}).
\paragraph{Full dipole dynamics.}
Combining Eqs.~\eqref{eq:self-propulsion}, \eqref{eq:external} and \eqref{eq:alpha}, the complete dynamical system for $N$ dipoles on the catenoid is
\begin{equation}
\begin{aligned}
\dot u_n &= \dot u_n^{(\text{self})} + \dot u_n^{(\text{ext})},\\[3pt]
\dot v_n &= \dot v_n^{(\text{self})} + \dot v_n^{(\text{ext})},\\[3pt]
\dot\alpha_n &= \dot\alpha_n^{(\text{self})} + \dot\alpha_n^{(\text{ext})}
+\tanh\!\big(\tfrac{v_n}{a}\big)\,\dot u_n.
\end{aligned}
\label{eq:main-system}
\end{equation}
To summarize the origin of the different contributions,
\begin{itemize}
\item $\dot u_n^{(\text{self})}, \dot v_n^{(\text{self})},\dot\alpha_n^{(\text{self})}$ are the curvature--corrected self--propulsion terms from Eq.~\eqref{eq:self-propulsion};
\item $\dot u_n^{(\text{ext})}, \dot v_n^{(\text{ext})},\dot\alpha_n^{(\text{ext})}$ are the external interaction terms from Eq.~\eqref{eq:external};
\item the final term in the orientation evolution represents the parallel--transport correction to the rotational velocity on the surface.
\end{itemize}

Figure~\ref{fig:dpsystem_geodesic} provides a numerical validation of the finite--dipole system derived above. 
For this test we consider the single--dipole case ($N=1$), so that the external interaction terms 
$\dot u_n^{(\mathrm{ext})}$, $\dot v_n^{(\mathrm{ext})}$ and $\dot\alpha_n^{(\mathrm{ext})}$ vanish identically, 
keeping only the leading self--propulsion terms of Eq.~(\ref{selfp_truncated}). The dipole is initialized such that $\Lambda>1$ so that we are in the supercritical class of trapped one-sided geodesics. In panel (a) we compare the time evolution of the dipole centre $(u(t),v(t))$ obtained from the full self--propelled dipole equations \eqref{eq:main-system} (blue curves) with the corresponding catenoid geodesic constructed from the same initial data (green curves). The two trajectories are almost indistinguishable over the entire integration interval, the geodesic curve is intentionally truncated in time to make this overlap visible. The agreement improves further if we go beyond the truncated expressions of Eq.~(\ref{selfp_truncated} and use the full expressions from Appendix Sec.~(\ref{fullexp}). This agreement shows that, once the leading curvature--corrected self--propulsion terms \eqref{eq:self-propulsion} are included (as well as the curvature mediated orientation rate), the finite dipole indeed moves along the geodesic of the supercritical class. Panel (b) displays the   the geodesic conserved quantities, $ L(t)$ and $ E(t)$. Both errors stay below $10^{-7}$, confirming that the numerical scheme preserves the Lagrangian structure and that the key terms in \eqref{eq:main-system} (self-propulsion  and parallel transport contribution to orientation dynamics) are essential ingredients to build the finite-sized dipole system on the catenoid, and on any curved surface in general. 

\section{Conclusion}
\label{cncl}

In this work, we developed and tested a geometric description of vortex dipoles on a surface of variable negative curvature, taking the catenoid of throat radius $a$ as a canonical example. Starting from the catenoid metric and its induced symplectic structure, we derived the point–vortex Hamiltonian in closed form, including the curvature-dependent self term $-\frac{1}{4\pi}\Gamma_i^2 \log h(v_i)$. This allowed us to identify the natural $U(1)$ symmetry associated with azimuthal rotations and to construct the corresponding momentum map
\[
J = \sum_{i=1}^N \Gamma_i\left( \frac{a}{2}v_i + \frac{a^2}{4}\sinh\frac{2v_i}{a} \right),
\]
which we verified numerically to be conserved, together with the Hamiltonian, for arbitrary throat radius. These invariants then served as stringent diagnostics for all subsequent simulations in this work.

A central result of the paper is the explicit confirmation of Kimura’s geodesic conjecture on a nontrivial minimal surface. By specializing the Hamiltonian dynamics to a two-vortex system of opposite circulations and passing to mean variables $(\bar u,\bar v)$, we showed that tightly bound dipoles propagate along catenoid geodesics, with the single dimensionless parameter
\[
\Lambda = \frac{a^2 \cosh^2(\bar v/a)\,\dot{\bar u}}{a \sqrt{\cosh^2(\bar v/a)(\dot{\bar v}^2 + a^2 \dot{\bar u}^2)}}
\]
classifying the orbits into meridional ($\Lambda=0$), circular-neck ($|\Lambda|=1$), and trapped one-sided (supercritical) ($|\Lambda|>1$) families. Figures~\ref{figgdsc1}–\ref{figgdsc3} demonstrate that the dipole trajectories obtained from the full vortex equations lie on top of the corresponding geodesic solutions, and that the relative errors in both $H$ and $J$ remain at or below $10^{-7}$.

We then used the same Hamiltonian framework to study genuinely dynamical phenomena that do not appear in the infinitesimal dipole limit. In Sec.~\ref{scattering}, we showed that the catenoid supports both \emph{direct} and \emph{exchange} scattering of classical vortex dipoles: depending on a small asymmetry in the initial impact geometry, the colliding vortex pairs either retain their identity or exchange partners. The fact that $H$ and $J$ remain conserved throughout the entire collision confirms that the scattering is mediated purely by curvature and not by numerical artefacts. In Sec.~\ref{cluster}, we contrasted this with the case of same-sign vortices, where we found long-lived, collectively rotating configurations (co-rotating four-vortex states) with overall azimuthal drift, again accompanied by conservation of $H$ and $J$.

Finally, in Sec.~\ref{dipolesystem}, we constructed an explicit \emph{finite-dipole} dynamical system on the catenoid. By displacing the two vortices of a dipole by $\pm \ell_n/2$ along a tangent direction $\alpha_n$, inserting these positions into the vortex equations, and averaging, we obtained closed expressions for the curvature-corrected self-propulsion,
\[
\dot u_n^{(\text{self})} \sim \frac{\sech(v_n/a)\,\sin\alpha_n}{\ell_n}, 
\qquad
\dot v_n^{(\text{self})} \sim -\frac{\sech(v_n/a)\,\cos\alpha_n}{\ell_n},
\]
together with an evolution equation for the orientation that contains both shear-induced and parallel-transport terms. This provides a concrete realization, on a curved minimal surface, of the intuitive statement that a finite dipole \emph{propels} orthogonally to the direction of its axis, with a speed modulated by curvature. For convenience, a concise summary of the primary dipole motion regimes studied in this work, together with their controlling dimensionless parameters and dynamical level, is collected in Table~\ref{tab:modes_catenoid}.

There are several natural directions for future work. An immediate extension of the work presented here would be to study the scattering of a self-propelled dipole with a co-rotating cluster of vortices, which is also of interest in the context of BEC experiments. The dynamics of vortex clusters on surfaces of varying negative curvature will also be reported in future work, in line with Ref.~\cite{sam3}. 

Finally, while the present work establishes the geometric structure of single-dipole motion and demonstrates representative two-dipole scattering channels, a systematic exploration of the full two-dipole phase space in the finite-sized dipole model remains an important direction for future investigation. In particular, by fixing the curvature ratio $\ell/a$ (dipole size relative to throat radius) and scanning the reduced dimensionless parameters $(R_0/\ell,\Delta\alpha_0,\bar v_0/a)$—where $R_0$ denotes the initial geodesic separation between dipole centres, $\Delta\alpha_0$ their initial relative orientation, and $\bar v_0$ the initial mean meridional height—one may construct a complete regime diagram distinguishing free geodesic drift, direct and exchange scattering, curvature-trapped motion, and possible long-lived bound oscillatory states. Such a classification would provide a global dynamical portrait of interacting dipoles on negatively curved surfaces and clarify how curvature, through the scale $a$, acts as a tunable control parameter for dipole scattering and collective behaviour.

On the mathematical side, the current work can be generalized in several directions. First, the present analysis was carried out on the catenoid because of its analytic tractability; it would be interesting to repeat the construction on other minimal or negatively curved surfaces (e.g., pseudospherical patches or corrugated films) to determine how universal the self-propulsion terms are. Second, our finite-dipole model retains only the leading order in the separation $\ell$; incorporating higher-order corrections would provide a controlled way to quantify the generic behaviour of a large many-body system of dipoles on the catenoid, using our full expressions in Appendix~\ref{fullexp}. Third, the catenoid offers a natural setting for studying vortex interactions in curved BEC trap geometries, where the present formulation may be viewed as a simplified version of full Gross–Pitaevskii simulations. Finally, since the momentum map $J$ is explicit, it may be used to attempt a more systematic classification of scattering channels on the catenoid, which we leave for future work.

In summary, our work shows that: (i) vortex dipoles on a catenoid follow geodesics classified by a single parameter $\Lambda$; (ii) the full point–vortex Hamiltonian conserves both $H$ and the azimuthal momentum map $J$ for arbitrary throat radius; (iii) surfaces of varying negative curvature such as the catenoid admit direct and exchange scattering of dipoles; and (iv) a simple finite-sized dipole model yields curvature-corrected self-propulsion and orientation dynamics, explicitly validated numerically. We hope this work will motivate further analytical and numerical studies of vortex matter on curved manifolds, where geometry can serve as a tunable control parameter for vortex transport.

\bmhead{Acknowledgements}
We are very thankful to Suryateja Gavva, Naomi Oppenheimer and Haim Diamant. R.S is supported by DST INSPIRE Faculty fellowship, India (Grant No.IFA19-PH231). Both authors acknowledge support from NFSG and OPERA Research Grant from Birla Institute of Technology and Science, Pilani (Hyderabad Campus). K.B. thanks the  Undergraduate Summer Program 2026 at the Nordic Institute for Theoretical Physics (Nordita).

\section*{Declarations}
All data generated or analyzed during this study are included in this  article. The computational codes used in this work are available from the corresponding author upon reasonable request.
\begin{appendices}

\section{Solution of the geodesic orbit integrals}
\label{appgeo}
In this appendix we solve explicitly the first–order ``orbit" equation for geodesic trajectories on the catenoid
\begin{equation}
\label{eq:traj-ode}
\frac{du}{dv}
= \pm \frac{\Lambda}{a\,\sqrt{\cosh^{2}\!\left(\frac{v}{a}\right) - \Lambda^{2}}}\,,
\end{equation}
where $a>0$ is the catenoid throat radius and $\Lambda$ is a constant fixed by the conserved quantities defined in the main text. The sign $\pm$ corresponds to the two orientation choices of the curve. The structure of the solution depends on whether $|\Lambda|>1$ or $|\Lambda|<1$, so we treat these cases separately.
Throughout, $F(\phi|m)$ denotes the incomplete elliptic integral of the first kind,
\begin{equation}
F(\phi|m) := \int_{0}^{\phi} \frac{d\theta}{\sqrt{1 - m \sin^{2}\theta}}\,.\nn
\label{eq:ellipticF-def}
\end{equation}
We first remove the length scale $a$ from the independent variable via
\begin{equation}
x := \frac{v}{a} \qquad \Longrightarrow \qquad dv = a\,dx.\nn
\end{equation}
Then \eqref{eq:traj-ode} becomes
\begin{equation}
\label{eq:traj-ode-x}
\frac{du}{dx}
= \pm \frac{\Lambda}{\sqrt{\cosh^{2}x - \Lambda^{2}}}\,,\nn
\end{equation}
and the problem reduces to the single integral
\begin{equation}
u(x) - u_{0} = \pm \Lambda \int \frac{dx}{\sqrt{\cosh^{2}x - \Lambda^{2}}}\,,
\label{eq:master-int}
\end{equation}
where $u_{0}$ is an integration constant. We now evaluate \eqref{eq:master-int} in the two regimes, such that we can bring the integrals to standard elliptic integral forms in the respective regimes.

\subsection*{\texorpdfstring{$|\Lambda|>1$}{|Lambda|>1} (supercritical case)}

When $|\Lambda|>1$ we write
\begin{equation}
\cosh^{2}x - \Lambda^{2} 
= \sinh^{2}x - (\Lambda^{2}-1).\nn
\end{equation}
Let us introduce
\begin{equation}
\alpha^{2} := \Lambda^{2} - 1 > 0.\nn
\end{equation}
Then
\begin{equation}
u - u_{0}
= \pm \Lambda \int \frac{dx}{\sqrt{\sinh^{2}x - \alpha^{2}}}.
\label{eq:int-L>1}\nn
\end{equation}
Set $z = \sinh x$. Then
\begin{equation}
u - u_{0} = \pm \Lambda \int \frac{dz}{\sqrt{1+z^{2}}\,\sqrt{z^{2}-\alpha^{2}}}.\nn
\end{equation}
We now introduce
\begin{equation}
z = \frac{\alpha}{\sqrt{1-y^{2}}}, \qquad |y|<1.\nn
\end{equation}
such that  the integral becomes
\begin{equation}
u - u_{0}
= \pm \Lambda \int \frac{dy}{\sqrt{1-y^{2}}\,\sqrt{\Lambda^{2}-y^{2}}}.\nn
\end{equation}
We set $y = \sin\phi$. Then
\begin{equation}
u - u_{0}
= \pm \int \frac{d\phi}{\sqrt{1 - \frac{1}{\Lambda^{2}}\sin^{2}\phi}}
= \pm F\!\left(\phi \,\bigg|\, \frac{1}{\Lambda^{2}}\right).\nn
\end{equation}
 We express $\phi$ in terms of the original variable $v$. Tracing back the substitutions,
the solution can be written as
\begin{equation}
u(v) - u_{0}
= \pm
F\!\left(
\arcsin \frac{\sqrt{\cosh^{2}\!\left(\tfrac{v}{a}\right) - \Lambda^{2}}}
               {\sinh\!\left(\tfrac{v}{a}\right)}
\;\middle|\;
\frac{1}{\Lambda^{2}}
\right), \qquad |\Lambda|>1.
\label{eq:sol-L>1}
\end{equation}
For $|\Lambda|>1$ the radicand $\cosh^{2}(v/a) - \Lambda^{2}$ is non–negative only if
\begin{equation}
\cosh\!\left(\frac{v}{a}\right) \ge |\Lambda|
\quad\Longrightarrow\quad
|v| \ge v_{\text{tp}} := a\,\operatorname{arcosh}|\Lambda|.\nn
\end{equation}
Thus the trajectory has a turning point at $v = \pm v_{\text{tp}}$, as expected for this regime.

\subsection*{\texorpdfstring{$|\Lambda|<1$}{|Lambda|<1} (subcritical case)}

When $|\Lambda|<1$ we instead  use
\begin{equation}
\cosh^{2}x - \Lambda^{2}
= \sinh^{2}x + (1-\Lambda^{2}).\nn
\end{equation}
Let us now define
\begin{equation}
\beta^{2} := 1 - \Lambda^{2} \in (0,1),\nn
\end{equation}
so that \eqref{eq:master-int} becomes
\begin{equation}
u - u_{0}
= \pm \Lambda \int \frac{dx}{\sqrt{\sinh^{2}x + \beta^{2}}}.\nn
\label{eq:int-L<1}
\end{equation}
We change variables
\begin{equation}
\sinh x = \beta \sinh t.\nn
\end{equation}
Hence
\begin{equation}
u - u_{0}
= \pm \Lambda \int \frac{dt}{\sqrt{1 + \beta^{2} \sinh^{2} t}}.\nn
\end{equation}
Let $y = \tanh t$ such that we have
\begin{equation}
u - u_{0}
= \pm \Lambda \int \frac{dy}{\sqrt{1-y^{2}}\,\sqrt{1 - \Lambda^{2}y^{2}}}.\nn
\end{equation}
Finally we set $y = \sin\phi$ again. Then
\begin{equation}
u - u_{0}
= \pm \Lambda \int \frac{d\phi}{\sqrt{1 - \Lambda^{2}\sin^{2}\phi}}
= \pm \Lambda\, F(\phi | \Lambda^{2}).\nn
\end{equation}
Similar to the supercritical case, we express $\phi$ in terms of $v$, leading us to
\begin{equation}
u(v) - u_{0}
= \pm \Lambda\,
F\!\left(
\arcsin
\frac{\sinh\!\left(\tfrac{v}{a}\right)}
     {\sqrt{\cosh^{2}\!\left(\tfrac{v}{a}\right) - \Lambda^{2}}}
\;\middle|\;
\Lambda^{2}
\right),
\qquad |\Lambda|<1.
\label{eq:sol-L<1}
\end{equation}
For $|\Lambda|<1$ we have
\begin{equation}
\cosh^{2}\!\left(\frac{v}{a}\right) - \Lambda^{2} > 0
\quad \text{for all } v \in \mathbb{R},\nn
\end{equation}
so, unlike the $|\Lambda|>1$ case, there is no turning point and the solution is real for all $v$. Moreover,
\begin{equation}
\frac{\sinh^{2}\!\left(\tfrac{v}{a}\right)}
     {\cosh^{2}\!\left(\tfrac{v}{a}\right) - \Lambda^{2}}
\;\le\;
\frac{\sinh^{2}\!\left(\tfrac{v}{a}\right)}
     {\cosh^{2}\!\left(\tfrac{v}{a}\right) - 1}
= 1,\nn
\end{equation}
so the argument of the $\arcsin(\cdot)$ in \eqref{eq:sol-L<1} indeed lies in $[-1,1]$ for all $v$.
\textit{Mathematica} may return for \eqref{eq:int-L<1} the equivalent form
\begin{equation}
u(v) - u_{0}
= \pm \frac{i \Lambda}{\sqrt{1-\Lambda^{2}}}\,
\text{EllipticF}\!\left(\frac{i v}{a}, \frac{1}{1-\Lambda^{2}}\right),\nn
\end{equation}
which is related to \eqref{eq:sol-L<1} by the standard imaginary–argument identity for the elliptic integral. We have preferred the manifestly real expressions \eqref{eq:sol-L>1} and \eqref{eq:sol-L<1}.

\section{Ordered tight--dipole limit of $\Lambda$}
\label{app:LambdaFullDerivation}

In this appendix we derive the ordered tight--dipole limit of the
geodesic parameter $\Lambda$, starting directly from the exact
two--vortex equations on the catenoid. The purpose is to make explicit
the limiting procedure $\Delta u\to0$ followed by $\Delta v\to0$
used in the main text.

We consider two vortices with circulations
$\Gamma_1=+1$ and $\Gamma_2=-1$, governed by
Eq.~(\ref{dyneq}). For $N=2$ the system reads
\begin{align}
\dot v_1 &= \frac{1}{4\pi a h^2(v_1)}
\frac{\Gamma_2 \sin(u_1-u_2)}{F_{12}}, \nn\\
\dot v_2 &= \frac{1}{4\pi a h^2(v_2)}
\frac{\Gamma_1 \sin(u_2-u_1)}{F_{21}}, \nn \\
\dot u_1 &= -\frac{1}{4\pi a^2 h^2(v_1)}
\frac{\Gamma_2 \sinh((v_1-v_2)/a)}{F_{12}}
+\frac{1}{4\pi a^2 h^2(v_1)}\Gamma_1 \tanh(v_1/a),\nn \\
\dot u_2 &= -\frac{1}{4\pi a^2 h^2(v_2)}
\frac{\Gamma_1 \sinh((v_2-v_1)/a)}{F_{21}}
+\frac{1}{4\pi a^2 h^2(v_2)}\Gamma_2 \tanh(v_2/a), \nn
\end{align}
where
\[
F_{12}=
\cosh\!\left(\frac{v_1-v_2}{a}\right)
-
\cos(u_1-u_2),
\qquad
h(v)=\cosh(v/a).
\]
Introducing the mean and separation variables
\[
\bar u=\frac{u_1+u_2}{2}, \quad
\Delta u=u_1-u_2,
\qquad
\bar v=\frac{v_1+v_2}{2}, \quad
\Delta v=v_1-v_2,
\]
we have
\[
u_{1,2}=\bar u\pm\frac{\Delta u}{2},
\qquad
v_{1,2}=\bar v\pm\frac{\Delta v}{2}.
\]
The mean velocities are
\[
\dot{\bar u}=\frac{\dot u_1+\dot u_2}{2},
\qquad
\dot{\bar v}=\frac{\dot v_1+\dot v_2}{2}.
\]

The geodesic parameter is defined as
\begin{equation}
\Lambda
=
\cosh\!\left(\frac{\bar v}{a}\right)
\frac{a\,\dot{\bar u}}
{\sqrt{\dot{\bar v}^{\,2}+a^2\dot{\bar u}^{\,2}}}.
\label{eq:LambdaDefApp}
\end{equation}

We now evaluate this quantity for small dipole size.
For $(\Delta u,\Delta v)\ll1$,
\begin{align}
\sin(\Delta u)&=\Delta u+O(\Delta^3), \qquad
\cos(\Delta u)=1-\tfrac12(\Delta u)^2+O(\Delta^4), \nonumber\\
\sinh(\Delta v/a)&=\tfrac{\Delta v}{a}+O(\Delta^3), \qquad
\cosh(\Delta v/a)=1+\tfrac12\!\left(\tfrac{\Delta v}{a}\right)^2+O(\Delta^4).\nn
\end{align}
Hence
\begin{equation}
F_{12}
=
\frac{1}{2}\left(\frac{\Delta v}{a}\right)^2
+
\frac{1}{2}(\Delta u)^2
+O(\Delta^4).
\label{eq:FExpansion}
\end{equation}
To leading order one may replace
$h(v_1)\simeq h(v_2)\simeq h(\bar v)$. Substituting these expansions gives
\begin{align}
\dot{\bar v}
&=
\frac{1}{4\pi a h^2(\bar v)}
\frac{\Delta u}{F_{12}}
+O(\Delta), \nn\\
\dot{\bar u}
&=
-\frac{1}{4\pi a^2 h^2(\bar v)}
\frac{\Delta v/a}{F_{12}}
+O(\Delta).\nn
\end{align}
Using \eqref{eq:FExpansion} yields
\begin{align}
\dot{\bar v}
&=
\frac{2a\,\Delta u}
{4\pi h^2(\bar v)
\left(\Delta v^2+a^2\Delta u^2\right)}
+O(\Delta), \\
\dot{\bar u}
&=
-\frac{2\,\Delta v}
{4\pi a h^2(\bar v)
\left(\Delta v^2+a^2\Delta u^2\right)}
+O(\Delta).
\end{align}
We now implement the ordered tight--dipole limit
\[
\Delta u\to0 \quad \text{followed by} \quad \Delta v\to0.
\]
In this ordering $\dot{\bar v}\to0$, while
\[
\dot{\bar u}
\sim
-\frac{2}{4\pi a h^2(\bar v)}
\frac{1}{\Delta v}.
\]
Consequently,
\[
\sqrt{\dot{\bar v}^2+a^2\dot{\bar u}^2}
=
\frac{2}{4\pi h^2(\bar v)}
\frac{1}{|\Delta v|}.
\]
Substituting into \eqref{eq:LambdaDefApp} gives
\begin{align}
\Lambda
&=
\cosh\!\left(\frac{\bar v}{a}\right)
\frac{a\,\dot{\bar u}}
{\sqrt{\dot{\bar v}^2+a^2\dot{\bar u}^2}} \nn\\
&=
\cosh\!\left(\frac{\bar v}{a}\right)
\frac{\Delta v}{|\Delta v|}
+
O(\text{dipole size}^2).\nn
\end{align}
Thus, in the tight--dipole limit,
\[
\Lambda
=
\cosh\!\left(\frac{\bar v}{a}\right)
\,\mathrm{sgn}(\Delta v)
+
O(\text{dipole size}^2).
\]
The magnitude of $\Lambda$ is determined by the local metric factor
$\cosh(\bar v/a)$, while its sign encodes the orientation of the dipole.
The appearance of $\mathrm{sgn}(\Delta v)$ originates from the absolute
value in the normalization
$\sqrt{\dot{\bar v}^2+a^2\dot{\bar u}^2}$.
For completeness, we record the exact intermediate expression
obtained after expanding $\Lambda$ first in $\Delta u$
and subsequently in $\Delta v$, prior to taking the ordered limit.
After simplification one finds
\begin{align}
\Lambda
&=
-\frac{\Delta u^2 \Delta v
\left(
435\tanh^4\!\left(\frac{\bar v}{a}\right)
-540\tanh^2\!\left(\frac{\bar v}{a}\right)
+121
\right)
\sech\!\left(\frac{\bar v}{a}\right)}
{480 a^2
\sqrt{\dfrac{\sech^4\!\left(\frac{\bar v}{a}\right)}{\Delta v^2}}}
\nonumber\\
&\quad
-\frac{a^2 \Delta u^2
\sech\!\left(\frac{\bar v}{a}\right)}
{2 \Delta v^3
\sqrt{\dfrac{\sech^4\!\left(\frac{\bar v}{a}\right)}{\Delta v^2}}}
\nonumber\\
&\quad
-\frac{
\sech\!\left(\frac{\bar v}{a}\right)
\left(
9\Delta u^2\tanh^2\!\left(\frac{\bar v}{a}\right)
-5\Delta u^2-12
\right)
}
{12 \Delta v
\sqrt{\dfrac{\sech^4\!\left(\frac{\bar v}{a}\right)}{\Delta v^2}}}
+ O(\Delta^2).
\label{eq:LambdaIntermediateCAS}
\end{align}
Using
\[
\sqrt{\frac{\sech^4(\bar v/a)}{\Delta v^2}}
=
\frac{\sech^2(\bar v/a)}{|\Delta v|},
\]
and implementing the ordered limit $\Delta u\to0$ followed by $\Delta v\to0$,
all terms proportional to $\Delta u^2$ vanish and the surviving leading contribution reduces to
\[
\Lambda
=
\cosh\!\left(\frac{\bar v}{a}\right)
\frac{\Delta v}{|\Delta v|}
+
O(\text{dipole size}^2).
\]

\section{Orientation dynamics}
\label{app:orientation}
We derive here the evolution equation governing the orientation of a finite vortex dipole on a curved surface, with the catenoid as our primary example. Let the right- and left-hand vortices of the dipole, with circulations \(+\Gamma\) and \(-\Gamma\), occupy positions
\begin{equation}
\mathbf{x}_+ = \mathbf{x}_d + \tfrac{1}{2}\mathbf{d}, 
\qquad
\mathbf{x}_- = \mathbf{x}_d - \tfrac{1}{2}\mathbf{d},
\end{equation}
where \(\mathbf{x}_d\) denotes the dipole center and \(\mathbf{d}\) is the separation vector lying in the local tangent plane. The dipole length is \(\ell = |\mathbf{d}|\), and its orientation is represented by the unit vector
\begin{equation}
\hat{\mathbf{e}}_{\parallel} = \frac{\mathbf{d}}{\ell},
\qquad
\hat{\mathbf{e}}_{\perp} = \hat{\mathbf{n}} \times \hat{\mathbf{e}}_{\parallel},
\end{equation}
where \(\hat{\mathbf{n}}\) is the unit normal to the surface. Differentiating the dipole vector gives
\begin{equation}
\dot{\mathbf{d}} = \dot{\mathbf{x}}_+ - \dot{\mathbf{x}}_- 
   = \dot{\ell}\,\hat{\mathbf{e}}_{\parallel}
     + \ell\,\dot{\hat{\mathbf{e}}}_{\parallel}.
\end{equation}
Under the finite-dipole constraint the separation magnitude remains constant (\(\dot{\ell}=0\)), so
\begin{equation}
\dot{\hat{\mathbf{e}}}_{\parallel}
   = \frac{1}{\ell}(\dot{\mathbf{x}}_+ - \dot{\mathbf{x}}_-)
   = \frac{1}{\ell}\,\Delta\mathbf{V},
\end{equation}
where \(\Delta\mathbf{V}\) is the relative velocity of the two vortices in the tangent plane. The orientation of the dipole is described by an angle \(\alpha\), measured from the local meridional (\(u\)) direction. 
By construction,
\begin{equation}
\dot{\hat{\mathbf{e}}}_{\parallel} = \dot{\alpha}\,\hat{\mathbf{e}}_{\perp}.
\end{equation}
Projecting the previous expression onto \(\hat{\mathbf{e}}_{\perp}\) yields the general law
\begin{equation}
\dot{\alpha}
   = \frac{1}{\ell}\,
      \hat{\mathbf{e}}_{\perp}\!\cdot\!
      (\dot{\mathbf{x}}_+ - \dot{\mathbf{x}}_-).
\label{eq:alpha_general}
\end{equation}
Equation~(\ref{eq:alpha_general}) states that the instantaneous angular velocity of the dipole axis is determined by the component of the differential vortex velocity perpendicular to the dipole separation.\\
We now compute the contribution from parallel transport to the rotation rate. We introduce the orthonormal coframe
\(\theta^{1}=a\,h(v)\,du\) and \(\theta^{2}=h(v)\,dv\),
so that \(ds^{2}=(\theta^{1})^{2}+(\theta^{2})^{2}\).
The Cartan structure equations
\(d\theta^{1}+\omega\wedge\theta^{2}=0\) and
\(d\theta^{2}-\omega\wedge\theta^{1}=0\)
determine the Levi–Civita connection one–form~$\omega$.
Since
\(d\theta^{1}=-a h'(v)\,du\wedge dv\) and \(d\theta^{2}=0\),
one finds
\(\omega=(a h'/h)\,du\).
Using \(h'(v)=a^{-1}\sinh(v/a)\) gives
\begin{equation}
\omega=\tanh\!\left(\tfrac{v}{a}\right)\,du.
\end{equation}
Evaluating this on a trajectory
\(X=\dot u\,\partial_{u}+\dot v\,\partial_{v}\)
yields the instantaneous rotation rate of a parallel–transported vector:
\begin{equation}
\omega(X)=\tanh\!\left(\tfrac{v}{a}\right)\,\dot u.
\label{eq:rotation-rate}
\end{equation}
Hence only motion around the azimuthal direction (\(u\)) generates
a local rotation of transported vectors, with magnitude set by
the geometric factor \(\tanh(v/a)\).
Consequently, Eq.~(\ref{eq:alpha_general}) reduces to
\begin{equation}
\dot{\alpha}
   = \frac{1}{\ell}\,
     \hat{\mathbf{e}}_{\perp}\!\cdot\!
     (\dot{\mathbf{x}}_+ - \dot{\mathbf{x}}_-)
     + \tanh\!\left(\frac{v}{a}\right)\dot{u}_d,
\label{eq:alpha_catenoid}
\end{equation}
which coincides with Eq.~(6.25) of the main text.
The additional term proportional to \(\tanh(v/a)\) represents the geometric rotation of the local basis vectors as the dipole center moves along the surface. Equations~(\ref{eq:alpha_general}) and~(\ref{eq:alpha_catenoid}) show that the dipole orientation evolves due to \textit{differential advection} between its two constituent vortices. The curvature of the catenoid introduces an effective torque that rotates the dipole axis even when the relative velocity vanishes in flat-space coordinates.
In the planar limit \(a \to \infty\), the curvature term disappears, and the expression reduces to the classical finite-dipole result of Aref~\cite{aref} and Saffman~\cite{saffman}.

\section{Full expressions of self-propulsion terms}
\label{fullexp}
Here we provide the full expressions for the self-propulsion velocity  and rotation rate as described in the main text.
\begin{align}
\dot u_i^{(\text{self})}&= -\frac{1}{8a^{2}\pi\Big(\cos\!\big(\tfrac{\ell_i\cos\alpha_i\,\sech(\tfrac{v_i}{a})}{a}\big)
  -\cosh\!\big(\tfrac{\ell_i\,\sech(\tfrac{v_i}{a})\sin\alpha_i}{a}\big)\Big)}\nn\\[6pt]
&\qquad\times\Bigg\{%
\mu_i\,\sech^{2}\!\Big(\tfrac{v_i-\tfrac12\ell_i\sech(\tfrac{v_i}{a})\sin\alpha_i}{a}\Big)\,
\sinh\!\Big(\tfrac{\ell_i\,\sech(\tfrac{v_i}{a})\sin\alpha_i}{a}\Big)\nn\\[4pt]
&\qquad\qquad\qquad\quad\times\Bigg[\cos\!\Big(\tfrac{\ell_i\cos\alpha_i\,\sech(\tfrac{v_i}{a})}{a}\Big)
-\cosh\!\Big(\tfrac{\ell_i\,\sech(\tfrac{v_i}{a})\sin\alpha_i}{a}\Big)
\tanh\!\Big(\tfrac{v_i-\tfrac12\ell_i\sech(\tfrac{v_i}{a})\sin\alpha_i}{a}\Big)\Bigg]\nn\\[6pt]
&\qquad\qquad\qquad\;+\;\mu_i\,\sech^{2}\!\Big(\tfrac{v_i+\tfrac12\ell_i\sech(\tfrac{v_i}{a})\sin\alpha_i}{a}\Big)\,
\sinh\!\Big(\tfrac{\ell_i\,\sech(\tfrac{v_i}{a})\sin\alpha_i}{a}\Big)\nn\\[4pt]
&\qquad\qquad\qquad\quad\times\Bigg[-\cos\!\Big(\tfrac{\ell_i\cos\alpha_i\,\sech(\tfrac{v_i}{a})}{a}\Big)
+\cosh\!\Big(\tfrac{\ell_i\,\sech(\tfrac{v_i}{a})\sin\alpha_i}{a}\Big)
\tanh\!\Big(\tfrac{v_i+\tfrac12\ell_i\sech(\tfrac{v_i}{a})\sin\alpha_i}{a}\Big)\Bigg]
\Bigg\}\nn\\[8pt]
\end{align}
\begin{align}
\dot v_i^{(\text{self})}&=\quad \; \frac{\mu_i\Big\{\sech^{2}\!\big(\tfrac{v_i-\tfrac12\ell_i\sech(\tfrac{v_i}{a})\sin\alpha_i}{a}\big)
+\sech^{2}\!\big(\tfrac{v_i+\tfrac12\ell_i\sech(\tfrac{v_i}{a})\sin\alpha_i}{a}\big)\Big\}\,
\sin\!\big(\tfrac{\ell_i\cos\alpha_i\,\sech(\tfrac{v_i}{a})}{a}\big)}{8a\pi\Big(\cos\!\big(\tfrac{\ell_i\cos\alpha_i\,\sech(\tfrac{v_i}{a})}{a}\big)
  -\cosh\!\big(\tfrac{\ell_i\,\sech(\tfrac{v_i}{a})\sin\alpha_i}{a}\big)\Big)}\nn\\[6pt]\nn
  \end{align}
  \begin{align}
\dot \alpha_i^{(\text{self})}&=\frac{1}{4a\ell_i\pi\Big(\cos\!\big(\tfrac{\ell_i\cos\alpha_i\;\sech(\tfrac{v_i}{a})}{a}\big)
  -\cosh\!\big(\tfrac{\ell_i\;\sech(\tfrac{v_i}{a})\sin\alpha_i}{a}\big)\Big)}\nn\\[6pt]
&\qquad\times\mu_i\Bigg\{\;
-\sech\!\Big(\frac{\,v_i-\tfrac{1}{2}\ell_i\;\sech(\tfrac{v_i}{a})\sin\alpha_i\,}{a}\Big)\nn\\[4pt]
&\qquad\qquad\qquad\qquad\qquad\cdot\Bigg[
\cos\alpha_i\;\sin\!\Big(\frac{\ell_i\cos\alpha_i\;\sech(\tfrac{v_i}{a})}{a}\Big)
+\sin\alpha_i\;\sinh\!\Big(\frac{\ell_i\;\sech(\tfrac{v_i}{a})\sin\alpha_i}{a}\Big)
\Bigg]\nn\\[8pt]
&\qquad\qquad\qquad\qquad\qquad+\;\Bigg(\cos\!\Big(\frac{\ell_i\cos\alpha_i\;\sech(\tfrac{v_i}{a})}{a}\Big)
-\cosh\!\Big(\frac{\ell_i\;\sech(\tfrac{v_i}{a})\sin\alpha_i}{a}\Big)\Bigg)\nn\\[4pt]
&\qquad\qquad\qquad\qquad\qquad\qquad\cdot\sin\alpha_i\;
\tanh\!\Big(\frac{\,v_i-\tfrac{1}{2}\ell_i\;\sech(\tfrac{v_i}{a})\sin\alpha_i\,}{a}\Big)\nn\\[8pt]
&\qquad\qquad\qquad\qquad\qquad+\;\sech\!\Big(\frac{\,v_i+\tfrac{1}{2}\ell_i\;\sech(\tfrac{v_i}{a})\sin\alpha_i\,}{a}\Big)\nn\\[4pt]
&\qquad\qquad\qquad\qquad\qquad\qquad\cdot\Bigg[
\cos\alpha_i\;\sin\!\Big(\frac{\ell_i\cos\alpha_i\;\sech(\tfrac{v_i}{a})}{a}\Big)
+\sin\alpha_i\;\sinh\!\Big(\frac{\ell_i\;\sech(\tfrac{v_i}{a})\sin\alpha_i}{a}\Big)
\Bigg]\nn\\[8pt]
&\qquad\qquad\qquad\qquad\qquad\qquad+\;\Bigg(-\cos\!\Big(\frac{\ell_i\cos\alpha_i\;\sech(\tfrac{v_i}{a})}{a}\Big)
+\cosh\!\Big(\frac{\ell_i\;\sech(\tfrac{v_i}{a})\sin\alpha_i}{a}\Big)\Bigg)\nn\\[4pt]
&\qquad\qquad\qquad\qquad\qquad\qquad\qquad\cdot\sin\alpha_i\;
\tanh\!\Big(\frac{\,v_i+\tfrac{1}{2}\ell_i\;\sech(\tfrac{v_i}{a})\sin\alpha_i\,}{a}\Big)
\Bigg\}.
\end{align}

\end{appendices}


\end{document}